\documentclass{article}

\usepackage{arxiv}

\usepackage[utf8]{inputenc}
\usepackage[T1]{fontenc}
\usepackage[english]{babel}
\usepackage{url}
\usepackage{booktabs}
\usepackage{amsfonts}
\usepackage{amssymb}
\usepackage{mathtools}
\usepackage{nicefrac}
\usepackage{microtype}
\usepackage{graphicx}
\usepackage[numbers]{natbib}
\usepackage{doi}
\usepackage{anyfontsize}
\usepackage{pifont,xcolor}
\usepackage{textpos}
\usepackage{ragged2e}
\usepackage{rotating}
\usepackage{array}
\usepackage{makecell}
\usepackage{siunitx}
\usepackage{algorithm}
\usepackage{caption}
\usepackage{colortbl}
\usepackage{circledsteps}
\usepackage{algpseudocode}
\usepackage[absl]{overpic}
\usepackage{hyperref}
\usepackage{cleveref}

\newcommand{\be}{\begin{equation}}
\newcommand{\ee}{\end{equation}}

\newcommand{\bfm}{\boldsymbol{m}}

\newcommand{\bfo}{\boldsymbol{0}}
\newcommand{\bfp}{\boldsymbol{p}}

\newcommand{\bfu}{\boldsymbol{u}}

\newcommand{\bfv}{\boldsymbol{v}}

\providecommand{\varGamma}{\Gamma}
\providecommand{\varDelta}{\Delta}
\providecommand{\varTheta}{\Theta}
\providecommand{\varLambda}{\Lambda}
\providecommand{\varXi}{\Xi}
\providecommand{\varPi}{\Pi}
\providecommand{\varSigma}{\Sigma}
\providecommand{\varUpsilon}{\Upsilon}
\providecommand{\varPhi}{\Phi}
\providecommand{\varPsi}{\Psi}
\let\Gamma\varGamma
\let\Delta\varDelta
\let\Theta\varTheta
\let\Lambda\varLambda
\let\Xi\varXi
\let\Pi\varPi
\let\Sigma\varSigma
\let\Upsilon\varUpsilon
\let\Phi\varPhi
\let\Psi\varPsi

\newcommand{\bftheta}{\boldsymbol{\theta}}
\newcommand{\bfbeta}{\boldsymbol{\beta}}

\usepackage{algpseudocode}
\usepackage[absl]{overpic}

\def\clw{\color{white}}
\definecolor{darkpastelgreen}{rgb}{0.01, 0.74, 0.24}
\definecolor{colsurface}{rgb}{0.059511,0.223228,0.708376}

 \definecolor{colw}{rgb}{1, 1,1}

\title{{BODIESReg: An open-source pipeline for registering 3D body scans using pose-aligned initialization}}

\author{{Vikash Chaurasia}\thanks{{Corresponding author: V.C.Chaurasia@tudelft.nl}} \\
\textit{{Department of Biomechanical Engineering, Faculty of Mechanical Engineering}} \\
\textit{{Delft University of Technology, Delft, The Netherlands}} \\
\And
{Judit Cueto Fernandez} \\
\textit{{Department of Biomechanical Engineering, Faculty of Mechanical Engineering}} \\
\textit{{Delft University of Technology, Delft, The Netherlands}} \\
\And
{J. Micah Prendergast} \\
\textit{{Cognitive Robotics, Faculty of Mechanical Engineering}} \\
\textit{{Delft University of Technology, Delft, The Netherlands}} \\
\And
{Eline van der Kruk} \\
\textit{{Department of Biomechanical Engineering, Faculty of Mechanical Engineering}} \\
\textit{{Delft University of Technology, Delft, The Netherlands}}
}

\renewcommand{\shorttitle}{{BODIESReg}}

\hypersetup{
pdftitle={BODIESReg: An open-source pipeline for registering 3D body scans using pose-aligned initialization},
pdfauthor={Vikash Chaurasia, Judit Cueto Fernandez, J. Micah Prendergast, Eline van der Kruk},
pdfkeywords={3D body scan registration, SMPL, parametric body model, inverse kinematics, pose-aligned initialization, Chamfer distance, anthropometric measurements, biomechanics},
}

\begin{document}
\maketitle

\begin{abstract}
Biomechanical models are used to quantify and optimize human movement in clinical rehabilitation, sports science, and occupational health. Personalizing these models requires accurate identification of anatomical landmarks and body segment parameters, which can be derived from 3D body surface scans. Registering these surfaces to a canonical template is essential for automated landmark detection and model scaling. However, fitting parametric body models to real-world 3D point clouds remains challenging: non-linear optimization can converge to suboptimal solutions when target poses deviate substantially from the default pose of a template. We present BODIESReg, an open-source registration pipeline that addresses this initialization problem by constructing a pose-aligned mesh before distance minimization begins. In automatic mode, BODIESReg can be used for end-to-end registration without user intervention and processes multiple scans in batch. We evaluated BODIESReg on two complementary datasets: CHI3D, a synthetic dataset containing complex human poses, and MorphoMotion, a set of real-world 3D optical scans. Automatic registration succeeded for $82.9\%$ of CHI3D scans and for all MorphoMotion scans. Among successful registrations, mean surface-fit error were below 10~mm across both datasets. For cases where automatic initialization fails, we provide interactive tools for manual pose correction and correspondence selection. BODIESReg supports large-scale registration of 3D body scans for biomechanical research.
\end{abstract}


\section{Introduction} 
  \label{sec:intro}
Biomechanical models of the human body are widely used to quantify and optimize movement in rehabilitation, sports science, and occupational health. These models represent the body as a system of rigid segments, and musculoskeletal models extend this by incorporating actuators that represent muscle-tendon units, enabling estimation of joint forces, moments, and muscle activations that cannot be measured directly in vivo. The validity and utility of such models depend on how well they reflect the morphology of the individual being studied. Personalizing a generic musculoskeletal model requires subject-specific estimates of both the inertial body segment parameters and the underlying skeletal geometry, including joint center locations and joint axes. Subject-specific scaling, therefore, requires accurate morphological measurement.
 
Traditionally, model personalization has relied on manual measurements through palpation and marker placement to identify the anatomical landmarks that define joint centers, axes, and segment boundaries. Three-dimensional (3D) surface representations of the body obtained through optical scanning, magnetic resonance imaging (MRI), or computed tomography offer far richer morphological detail and enable more precise, automated extraction of these landmarks and segment parameters. Such surfaces are typically represented as a mesh of vertices (points in 3D space) and faces (triangles connecting those points), but in this raw form the data carry no anatomical meaning. To assign anatomical meaning to a scan, registration is used to establish correspondence between the acquired surface and a canonical template. With that correspondence in place, anatomical labels from the template can be transferred to the scan surface, enabling automatic, consistent extraction of anatomical landmarks across subjects, regardless of individual differences in size, shape, or pose. Developing reliable methods for this registration step and integrating them into accessible tools for the biomechanics community remains an open challenge.

Non-parametric registration methods deform a template mesh to match a target shape by
manually establishing a few correspondences between the template and target, then applying
deformation models such as As-Rigid-As-Possible (ARAP)~\cite{sorkine2007rigid} and
As-Conformal-As-Possible (ACAP)~\cite{yoshi2014acap}. Establishing accurate anatomical
correspondences is time-consuming and error-prone~\cite{makadia2006fully,amberg2007optimal},
usually requiring manual annotation or user interaction.
Ovsjanikov et al.~\cite{ovsjanikov2012functional} attempted to address this by using functional maps to represent
the correspondences in a spectral domain, allowing for softer matching between shapes. However,
classical functional-map pipelines still often rely on hand-crafted descriptors, careful
initialization, or supervision to recover reliable correspondences in practice
~\cite{ovsjanikov2012functional,sharma2020weakly,attaiki2021dpfm}. For biomechanical studies
with dozens or hundreds of subjects, manual correspondence steps make non-parametric pipelines
impractical as a primary registration tool.

Parametric methods register scans by optimizing the shape and pose parameters of a statistical body model to minimize geometric distance to the target. SCAPE~\cite{anguelov2005scape} separated pose and shape deformations statistically, and SMPL~\cite{loper2015smpl} introduced a differentiable, vertex-based representation suitable for gradient-based optimization. Later models increased expressiveness: SMPL-X~\cite{pavlakos2019expressive} added hands and face, and STAR~\cite{osman2020star} improved generalization with more diverse training data. However, registration is still commonly based on geometric distance, and distance-based objectives can produce geometrically close but anatomically incorrect registrations. This risk increases when initialization is poor: if the template pose differs substantially from the scan pose, closest-point correspondences can be anatomically wrong from the first iteration, and the optimizer can continue to reduce the distance objective while preserving these incorrect correspondences. Schmidt et al.~\cite{schmidt2024evaluation} reported that registration failures occurred when the target pose deviated substantially from the default T-pose template.

Several published methods produce SMPL parameter estimates but address a different input or objective. Image-based methods that fit SMPL to RGB images include SMPLify~\cite{bogo2016keep}, which recovers SMPL parameters from a single RGB image by minimizing reprojection error against 2D joint detections; SPIN~\cite{kolotouros2019learning}, which interleaves a learned regressor with iterative fitting to improve monocular reconstruction; and HybrIK~\cite{li2021hybrik}, which uses a hybrid analytical-neural inverse kinematics formulation that improves 3D pose accuracy on image benchmarks. These methods optimize consistency with 2D image projections rather than 3D scan geometry.

Among methods that estimate SMPL parameters from 3D point clouds or scans, LoopReg~\cite{bhatnagar2020loopreg} was evaluated using correspondence accuracy on synthetic dressed-body datasets and visual plausibility on clothed scans, rather than absolute surface-distance error against the underlying anatomical geometry. In addition, it relies on supervised learning, requiring annotations for approximately 1000 of the 2600 training samples. ArtEq~\cite{feng2023generalizing} proposed an articulated SE(3)-equivariant neural network that directly regresses SMPL pose and shape parameters from 3D point clouds, with an emphasis on improving zero-shot generalization to unseen poses. However, the absence of a publicly available pretrained model limits its use for automatic SMPL registration.
To the best of our knowledge, there is currently no published open-source framework that performs fully automatic registration of parametric body models to 3D surface scans without dataset-specific training.  

We developed BODIESReg, an open-source pipeline for registering parametric body models to 3D surface scans for biomechanical applications, with a target mean surface-fit error below 10~mm (\url{https://github.com/BODIES-Lab-TU-Delft/BODIESReg}, MIT license). BODIESReg automatically estimates scan pose from sparse keypoints, constructs a pose-aligned template, and performs distance minimization. Pose-aligned initialization reduces, but does not eliminate, the risk of anatomically incorrect registrations caused by poor initial correspondences. When severe noise or missing surface regions make automatic pose estimation unreliable, users can edit the pose or select correspondences using interactive tools.   

The manuscript is organized as follows. In Section~\ref{sec:preliminaries}, we introduce parametric body modeling and define the notation. In Section~\ref{sec:formulation}, we describe pose-aligned initialization, distance minimization, and correspondence refinement. In Section~\ref{sec:evaluation_error_analysis}, we evaluate BODIESReg on one synthetic dataset and one dataset of real optical body scans, including registration accuracy, failure cases, computational performance, and acceleration settings. In Section~\ref{sec:discussion}, we discuss the main
findings, design trade-offs, limitations, and future work.  
 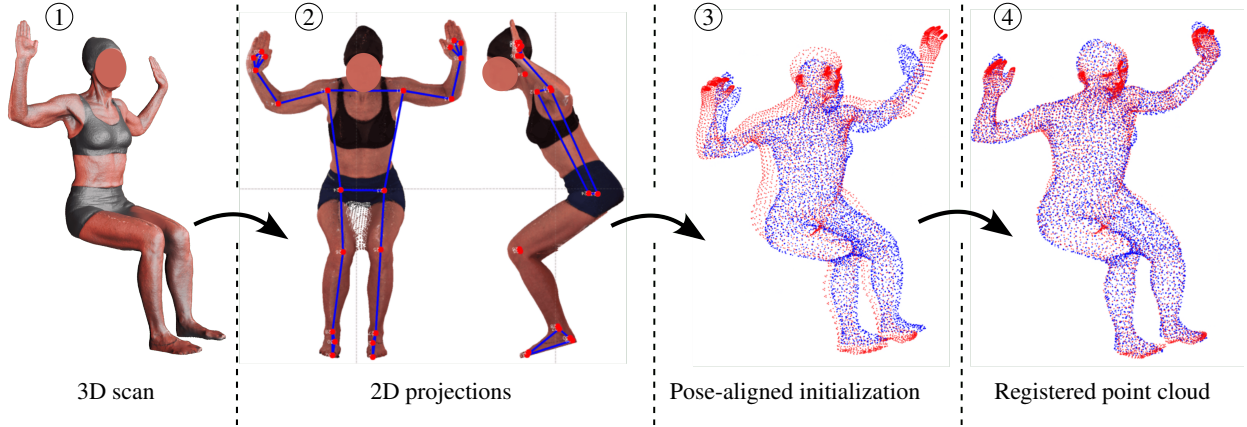
\begin{figure*}[ht!]
    \centering  
   \begin{overpic}[width=1\linewidth,  unit=1bp,tics=2 ] {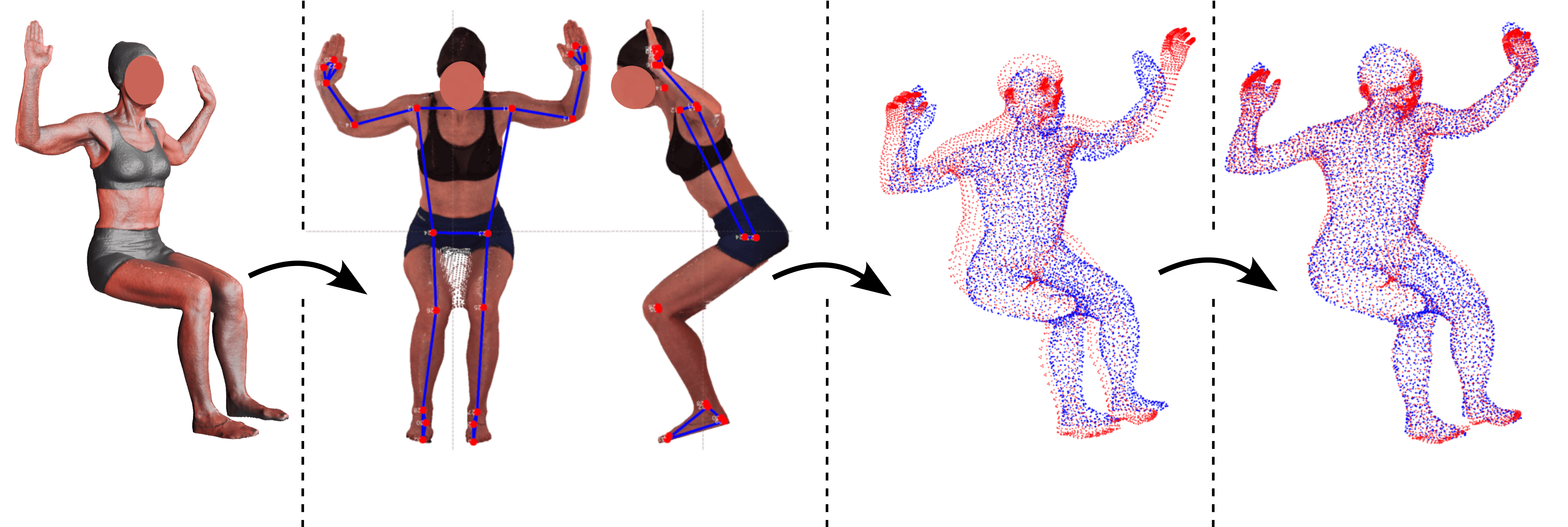}
\put(4,32){\small \Circled{1}}
 \put(6.5,2){\small 3D scan}
  \put(24,32){\small \Circled{2}}
  \put(30,2){\small  2D projections}
   \put(56,32){\small \Circled{3}}
      \put(80,32){\small \Circled{4}}
   \put(54,2){\small Pose-aligned  initialization}

    \put(80,2){\small Registered point cloud}
      
      \end{overpic}
\caption{Key steps of the BODIESReg registration pipeline.
\Circled{1} A representative 3D scan.
\Circled{2} Projections of the 3D scan into two orthogonal 2D views with keypoints
detected using MediaPipe.
\Circled{3} Pose-aligned initialization (red), superimposed on the 3D
scan (blue), obtained via inverse kinematics from the detected
keypoints.
\Circled{4} Registered point cloud (red) after pose and shape
optimization.}

        \label{fig:pipeline}
\end{figure*}
 \section{Preliminaries}
\label{sec:preliminaries}
 \subsection{Parametric Human Body Models}
\label{sec:param_model}
We describe a parametric human body model by a function $M$ that maps a low-dimensional set of parameters that govern shape and pose to the high-dimensional 3D vertex coordinates of a body scan:
\begin{align}\nonumber
    M: \mathbb{R}^{D_\beta} \times \mathbb{R}^{D_\theta} &\rightarrow \mathbb{R}^{N \times 3} \\
    (\bfbeta, \bftheta) &\mapsto M(\bfbeta, \bftheta), \label{eq:param_model_def}
\end{align}
where $M(\bfbeta, \bftheta)$ is an $N \times 3$ matrix representing the 3D positions of the $N$ vertices of the output mesh. The shape parameter vector $\bfbeta \in \mathbb{R}^{D_\beta}$ controls deformation of a base template shape to match a specific body type, accounting for variations in characteristics such as height, weight, and overall proportions. The pose vector $\bftheta \in \mathbb{R}^{D_\theta}$ represents the set of relative 3D rotations of the joints with respect to their parent joints in a kinematic tree structure, typically parameterized using methods such as axis-angle representations or quaternions.

In this work, we use the SMPL model~\cite{loper2015smpl} and its variants, which generate a 3D mesh $M$ with $N$ vertices from shape parameters $\bfbeta$ and pose parameters $\bftheta$. The pose is defined by axis-angle rotations for 23 anatomical joints plus a global root orientation, giving $D_\theta = 72$ pose parameters in total. SMPL parameterizes body shape using a low-dimensional PCA shape space and represents pose through linear blend skinning (LBS) driven by skeletal joint rotations. Before skinning, the template is deformed into $\mathbf{T}(\bfbeta, \bftheta)$ by adding shape-dependent displacements $B_S(\bfbeta)$ and pose-dependent blendshape correctives $B_P(\bftheta)$ to the mean template mesh $\bar{\mathbf{T}}$:
\begin{equation}
\mathbf{T}(\bfbeta, \bftheta) = \bar{\mathbf{T}} + B_S(\bfbeta) + B_P(\bftheta)
\end{equation}
 Here, $B_S(\bfbeta)$ is linear in $\bfbeta$ and encodes shape deviations from the template. $B_P(\bftheta)$ corrects for soft-tissue deformations induced by pose; it is linear in the per-joint rotation matrices $R(\bftheta)$, which are themselves nonlinear (trigonometric) functions of $\bftheta$. Rest-pose joint locations $J(\bfbeta)$ are subsequently computed using a linear regressor applied to the shape-deformed template vertices. The final posed mesh vertices $M( \bfbeta,\bftheta )$ are then obtained by applying the LBS function $W$ to the pre-skinning deformed template, driven by the joint locations, pose parameters, and predetermined skinning weights $\mathcal{W}$:
\begin{equation}
M(\bfbeta,\bftheta) = W(\mathbf{T}(\bfbeta,\bftheta), J(\bfbeta), \bftheta, \mathcal{W}).
\end{equation} 
%
%
\subsection{Forward and Inverse Kinematics}
\label{sec:fk_ik}
The pose parameters $\bftheta$ of the parametric model implicitly define the 3D locations of the joints through a kinematic chain. Given rest-pose joint locations $J(\bfbeta) = \{J_k\}_{k=1}^K$ and relative 3D rotations $R = \{R_{pa(k),k}\}_{k=1}^K$, the forward problem maps $R$ and $J(\bfbeta)$ to posed joint locations $Q = \text{FK}(R, J(\bfbeta))$ by traversing the kinematic tree from the root. Inverse kinematics (IK) reverses this: given a desired set of target 3D joint locations $P = \{p_k\}_{k=1}^K$, IK seeks $R$ such that $P = \text{FK}(R, J(\bfbeta))$. IK is generally ill-posed: target positions may be unreachable, or multiple rotation combinations can produce the same configuration (e.g., elbow-up vs.\ elbow-down for the same end-effector position).
\section{Registration Process}
\label{sec:formulation}
In Figure~\ref{fig:pipeline}, we show the key steps of the BODIESReg pipeline for registering a parametric model to a 3D body scan. We detect sparse 3D joint locations, estimate body pose through inverse kinematics, and instantiate a pose-aligned initial mesh. We then refine its shape and pose by minimizing the distance to the scan. The details of the pipeline are described in the following subsections.
\subsection{Pose Initialization}
\label{sec:pose_init}
 Instead of deforming a template from a default T-pose, our method estimates the pose of the input 3D scan and produces a pose-aligned initial template that is already geometrically close to the target shape.  

 We align the input scan to a canonical coordinate system, generate orthogonal 2D projections, and detect sparse 3D anatomical keypoints from projections or interactive input. We then estimate pose parameters via inverse kinematics and instantiate the SMPL template with these parameters to produce a pose-aligned initial mesh. 
\subsubsection{Coordinate System Alignment}
\label{sec:coord_alignment}
 Input 3D body scans may be acquired in arbitrary coordinate frames and represented in different coordinate units. We first detect the input unit automatically, or use the unit specified by the user, and convert the scan coordinates to meters to match the SMPL model units. Automatic unit detection uses the full-body extent of the scan, while explicit unit selection bypasses this inference. 

 We initialize the template in a fixed canonical pose before aligning the scan. Specifically, we set the template translation to $\bfo$ and seed the root orientation with the axis--angle vector $\bftheta_{\mathrm{root}}=(0,\pi/\sqrt{2},\pi/\sqrt{2})$. This seed corresponds to a rotation of $\pi$ about the axis $(0,1/\sqrt{2},1/\sqrt{2})$, equivalently a rotation by $\pi/2$ about the $X$-axis followed by a rotation by $\pi$ about the global $Z$-axis. This canonical initialization gives a template frame in which the vertical body axis is aligned with the $Z$-axis, with the body facing the positive $Y$-direction. The $XZ$ and $YZ$ planes correspond to the frontal and sagittal planes, respectively. 
 After unit normalization, we translate the scan so that its centroid matches the centroid of the seeded template. We then use PCA to estimate the dominant directions of the scan and the template. The scan is rotated so that its first and second principal directions align with the corresponding template directions, while the third direction is fixed by their cross product. Because PCA directions are ambiguous up to sign, we resolve the remaining flip by comparing signed point distributions along the third template axis and applying a $180$-degree rotation when needed. 
 We record the unit-conversion factor, centroid translation, and rotation matrices used to map the input scan into the SMPL coordinate frame. Registration is then performed in this SMPL frame. After registration, we apply the inverse transformations, and any optional manual alignment correction, to recover the registered mesh and inferred joints in the original scan coordinate system and units. 
\subsubsection{Orthogonal Projection Generation}
\label{sec:projection_generation}
From the aligned mesh, we generate two orthogonal projections onto the coordinate planes. In the aligned coordinate system, the $X$-axis is medial-lateral, the $Y$-axis is anterior--posterior, and the $Z$-axis is superior-inferior. The body faces the positive $Y$-direction. We render the front view from the positive $Y$-direction toward the negative $Y$-direction, yielding a projection onto the $XZ$-plane. We render the side view from the positive $X$-direction toward the negative $X$-direction, yielding a projection onto the $YZ$-plane. The corresponding mappings are 
\begin{align}
\Pi_{\text{front}} &: \mathbb{R}^3 \rightarrow \mathbb{R}^2,
& (x,y,z) &\mapsto (x,z),\\
\Pi_{\text{side}} &: \mathbb{R}^3 \rightarrow \mathbb{R}^2,
& (x,y,z) &\mapsto (y,z).
\end{align}
We render the front and side projections as images $I_{\text{front}}$ and $I_{\text{side}}$, respectively, each with spatial resolution $h\times w$.
\subsubsection{3D Keypoint Estimation}
\label{sec:keypoint_estimation}
We first estimate 2D anatomical keypoints in two orthogonal projections of the aligned scan using MediaPipe~\cite{lugaresi2019mediapipe}. MediaPipe returns 33 landmarks and their visibility scores in each projected image. We then map the normalized image coordinates from the frontal and sagittal projections back to the aligned 3D scan frame.

For keypoint $i$, let $(\hat{u}_i^{\mathrm{view}},\hat{v}_i^{\mathrm{view}})\in[0,1]^2$ denote the normalized MediaPipe coordinates in a projected image. The horizontal coordinate $\hat{u}$ increases from left to right, whereas the vertical coordinate $\hat{v}$ increases from top to bottom. The frontal projection has horizontal bounds $[x_{\min},x_{\max}]$ and vertical bounds $[z_{\min},z_{\max}]$. The sagittal projection has horizontal bounds $[y_{\min},y_{\max}]$ and the same vertical bounds $[z_{\min},z_{\max}]$. The corresponding aligned-frame coordinates are recovered by inverse linear mapping:
\be
\begin{aligned}
x_i^{\mathrm{front}}
&=x_{\min}
+\hat{u}_i^{\mathrm{front}}(x_{\max}-x_{\min}),\\
z_i^{\mathrm{front}}
&=z_{\max}
-\hat{v}_i^{\mathrm{front}}(z_{\max}-z_{\min}),\\[2pt]
y_i^{\mathrm{side}}
&=y_{\min}
+\hat{u}_i^{\mathrm{side}}(y_{\max}-y_{\min}),\\
z_i^{\mathrm{side}}
&=z_{\max}
-\hat{v}_i^{\mathrm{side}}(z_{\max}-z_{\min}).
\end{aligned}
\ee
We reconstruct the 3D MediaPipe keypoint $\bfp_i=(x_i,y_i,z_i)$ by taking the lateral coordinate from the frontal view, the anterior--posterior coordinate from the sagittal view, and the vertical coordinate from the frontal view:
\be
x_i = x_i^{\mathrm{front}},\qquad
y_i = y_i^{\mathrm{side}},\qquad
z_i = z_i^{\mathrm{front}}.
\ee
Using the frontal view as the reference for the vertical coordinate avoids making $z_i$ depend on relative detector confidence across views. The two estimates of $z_i$ could instead be fused using visibility scores, but confidence-weighted fusion can become unstable when one view has noisy detections. We therefore use the frontal estimate for $z_i$ and use visibility scores to reject keypoints below the selected threshold.

Because BODIESReg uses SMPL models for registration, the reconstructed MediaPipe keypoints are converted to the SMPL skeleton before pose fitting. In BODIESReg, we use only 22 of the $K=24$ SMPL joints as sparse 3D anatomical keypoints ($D_k=22$) for pose initialization. We exclude the two hand joints because they are not detected reliably from the orthogonal mesh projections. The converted SMPL keypoints retain confidence scores derived from the MediaPipe visibility scores, and we record whether each keypoint was directly detected or derived from neighboring landmarks. Details of the MediaPipe-to-SMPL conversion are provided in Appendix~\ref{sec:appendix_keypoint_mapping}.

A limitation of MediaPipe is that in near-symmetric poses, such as A-pose, it can assign left and right landmarks incorrectly. We automatically correct these assignments using the anatomical heuristics described in Appendix~\ref{sec:keypoint_left_right_correction}.

\begin{figure}[ht!]
    \centering
   \begin{overpic}[width=0.5\linewidth,  unit=2bp,tics=7 ] {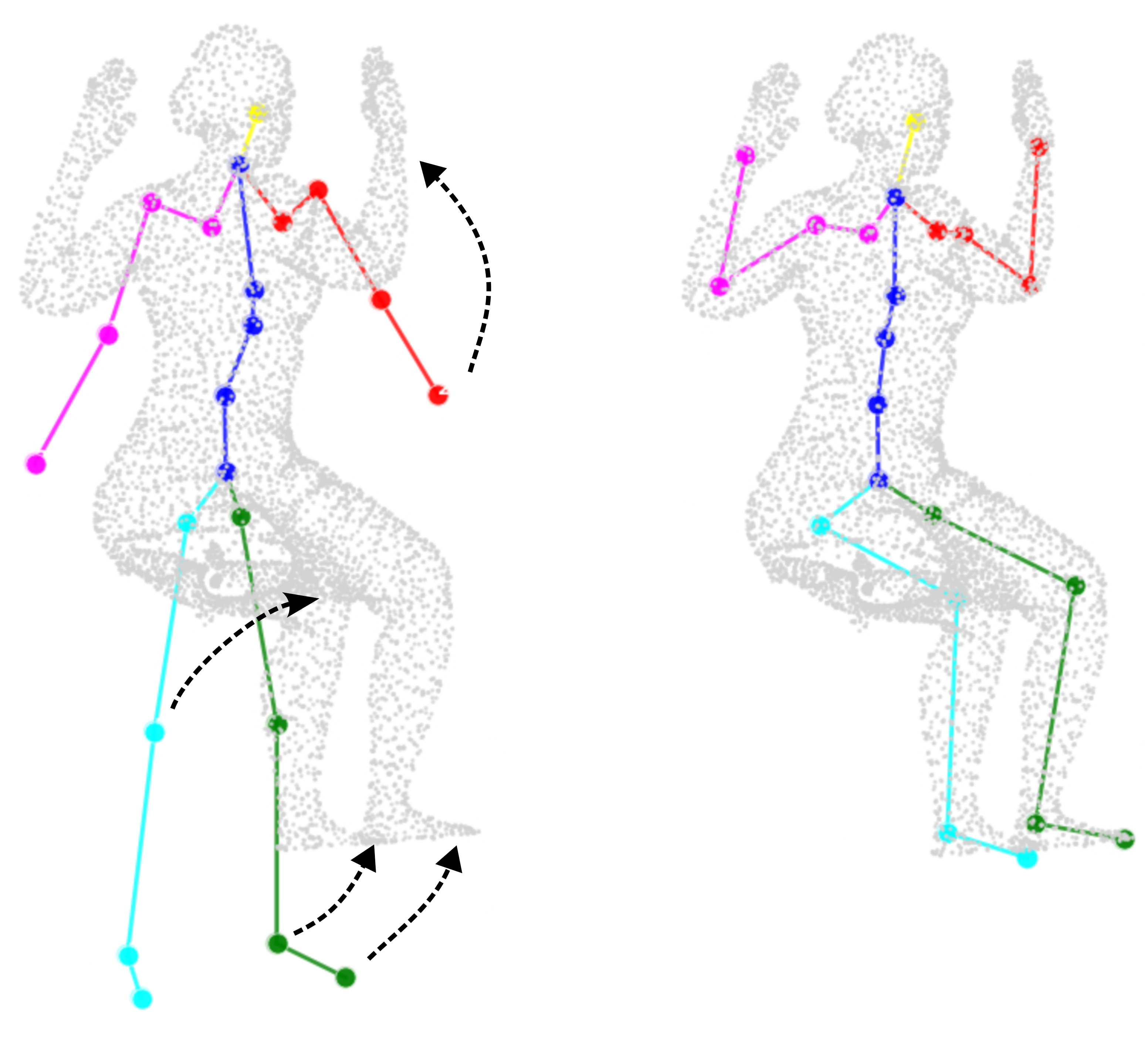}
    \end{overpic}
\caption{Interactive pose editor. Left: a 3D point cloud overlaid with the default 22-joint SMPL skeleton (two hand joints are excluded). Users drag joints to desired anatomical locations, as indicated by the arrows. Right: the updated skeleton after joint adjustment.}
        \label{fig:pose_editor}
\end{figure}

When automatic keypoint detection fails because of difficult poses, noise, missing color, or low detector confidence, users can position the 22 anatomical keypoints manually using the interactive pose editor shown in Figure~\ref{fig:pose_editor}. The editor displays the scan with a skeleton overlay and allows each keypoint to be moved toward its corresponding anatomical location. Because these manual placements are used only for pose initialization, approximate visual placement is sufficient; the subsequent registration optimization refines the body pose and surface fit.
\subsubsection{Inverse Kinematics with VPoser}
\label{sec:pose_estimation_initial_mesh}
We estimate pose parameters $\bftheta$ of the SMPL model by minimizing the distance between the SMPL joint locations $Q(\bfbeta,\bftheta)$ and the detected 3D keypoints $P=\{\mathbf{p}_i\}_{i=1}^{D_k}$. Optimizing directly over the full pose vector $\bftheta \in \mathbb{R}^{D_\theta}$ does not guarantee anatomically plausible results. Gradient descent on this high-dimensional, non-convex objective can converge to configurations where individual joints satisfy the distance criterion while the overall pose remains anatomically invalid. For example, the optimization may converge to a configuration with the head rotated by $180^\circ$ or bend a limb beyond its anatomical range.

To constrain the search to valid human poses, we adapt the VPoser inverse kinematics (IK) framework~\footnote{\url{https://github.com/nghorbani/human_body_prior}}, a
Variational Autoencoder (VAE) developed by Pavlakos et al.~\cite{pavlakos2019expressive}
and trained on motion-capture data. VPoser maps a low-dimensional latent code
$\mathbf{z} \in \mathbb{R}^{D_z}$ (with $D_z \ll D_\theta$) to a pose vector
$\bftheta = \text{Decoder}(\mathbf{z})$ that lies on the learned manifold of
valid human poses. We therefore optimize over $\mathbf{z}$ instead of
$\bftheta$ directly, minimizing:
\be\label{eq:pose_estimation_loss}
    E'(\bfbeta, \mathbf{z}) = \alpha\sum_{i=1}^{D_k} w_i \| Q_i(\bfbeta, \text{Decoder}(\mathbf{z})) - \mathbf{p}_i\|^2   + \lambda_z E_z(\mathbf{z}) + \lambda_\beta E_\beta(\bfbeta)
\ee
where $\alpha$ is the data term weight, $w_i \in [0,1]$ is the MediaPipe confidence score for keypoint $i$, $E_z(\mathbf{z}) = \|\mathbf{z}\|^2$ penalizes deviation from the VPoser prior (the latent space is trained as $\mathcal{N}(\mathbf{0},\mathbf{I})$, so this term acts as a Gaussian pose prior), and $E_\beta(\bfbeta) = \|\bfbeta\|^2$ penalizes deviation from neutral shape,  with weights $\lambda_z$ and $\lambda_\beta$.

 We introduce three modifications relative to the original VPoser IK formulation. First, the data term is weighted by $w_i$: joints with low detection confidence (for example, those occluded by the body or near the image boundary) contribute less to the objective, reducing their influence on the fit. Second, we empirically set the regularization weights to $\lambda_z = \lambda_\beta = 0.001$, substantially below the defaults of $0.01$ and $0.5$, respectively. A 3D body scan pose may differ substantially from a standard initialization, in which case strong regularization would prevent the optimizer from reaching the target joint configuration by penalizing deviations from the prior. Lower weights allow the optimizer to traverse the pose manifold more freely, while the VPoser decoder continues to enforce anatomical plausibility as a structural constraint. Third, we increase the data weight to $\alpha = 20$ (from $10$) to ensure that keypoint matching remains the dominant objective. 

Minimizing~\eqref{eq:pose_estimation_loss} yields parameters
$(\bfbeta_0, \bftheta_0)$ such that the joints of $M(\bfbeta_0, \bftheta_0)$
closely match the detected 3D keypoints.
 
The shape and pose parameters $(\bfbeta_0, \bftheta_0)$ are now instantiated in the SMPL model to generate a pose-aligned initial template mesh $M(\bfbeta_0, \bftheta_0)$. Unlike a default T-pose template that must be globally deformed to match the input pose, this mesh already exhibits a similar pose to the 3D body scan. This geometric proximity reduces the need for pose or shape regularization in the subsequent stages. The pose-aligned template is used as the initial condition for the distance minimization step described in the next subsection.
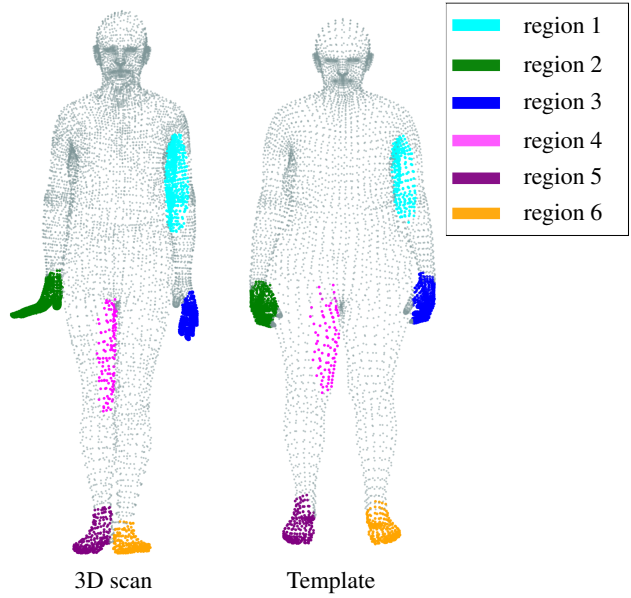
\begin{figure}[ht!]
    \centering  
   \begin{overpic}[width=0.5\linewidth,  unit=2bp,tics=4 ] {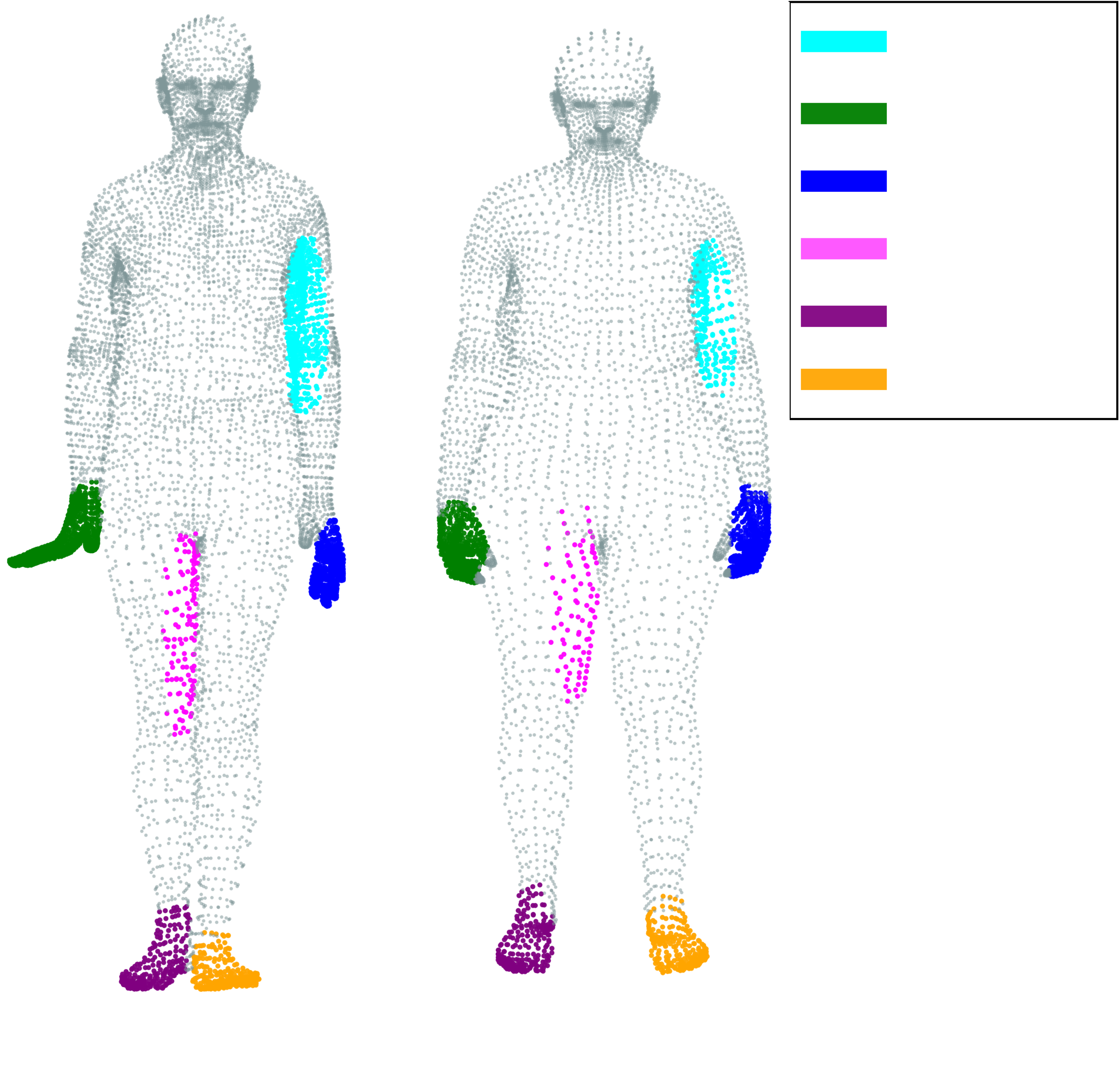}
   \put(11,2){\small 3D scan}
  \put(45,2){\small Template}
  \put(83,91){\small region 1}
  \put(83,84.8){\small region 2}
  \put(83,78.8){\small region 3}
  \put(83,72.5){\small region 4}
  \put(83,66.8){\small region 5}
 \put(83,61){\small region 6}
   \end{overpic}
\caption{Interactive correspondence selector used in manual mode. The user selects paired regions on the input 3D scan and template, shown in matching colors. During full-body optimization, Chamfer matching within each selected region is restricted to its paired region. Remaining unselected vertices are matched using the standard Chamfer objective.
} 
\label{fig:correspondence_selector}
\end{figure}
\subsection{Distance Minimization}
\label{sec:dist_refinement}
The pose-initialization stage provides an initial estimate $(\bfbeta_0,\bftheta_0)$ that places the SMPL mesh in approximate alignment with the scan. Starting from this estimate, we refine the shape and pose parameters by minimizing the geometric distance between the SMPL mesh $M(\bfbeta,\bftheta)$ and the aligned scan $S$. The optimized parameters $(\bfbeta^*,\bftheta^*)$ define the registered body shape and pose.

At this stage, point correspondences between $M$ and $S$ are unknown. We therefore measure surface proximity using the symmetric Chamfer distance, introduced by Barrow et al.~\cite{barrow1977parametric}, which does not require predefined correspondences. Let $\mathcal{V}(\bfbeta,\bftheta)=\{\mathbf{v}_j\}_{j=1}^N$ denote the vertices of the SMPL mesh and let $\mathcal{S}=\{\mathbf{u}_i\}_{i=1}^{N_S}$ denote the scan points. We minimize
\be 
    \tilde{E}_{\text{chamfer}}(\bfbeta,\bftheta)\\
    = \frac{1}{|\mathcal{V}|} \sum_{\mathbf{v} \in \mathcal{V}}
    \min_{\mathbf{u} \in \mathcal{S}} \| \mathbf{v} - \mathbf{u} \|
    + \frac{1}{|\mathcal{S}|} \sum_{\mathbf{u} \in \mathcal{S}}
    \min_{\mathbf{v} \in \mathcal{V}} \| \mathbf{u} - \mathbf{v} \| 
    + \lambda_\beta \|\bfbeta - \bfbeta_0\|^2
    + \lambda_\theta \|\bftheta - \bftheta_0\|^2.
    \label{eq:chamfer_opt}
\ee
initialized at $(\bfbeta_0,\bftheta_0)$ from the pose-initialization stage. 
 The first two terms enforce bidirectional surface proximity without requiring pre-established correspondences: the first penalizes model vertices with no nearby scan point, and the second penalizes scan points with no nearby model vertex. The last two terms are regularizers that anchor the parameters to the inverse-kinematics estimates and reduce drift in regions with sparse scan coverage. We used weak regularization weights, $\lambda_\beta = \lambda_\theta = 0.001$, for all experiments. These weights keep regularization small relative to the geometric fitting objective while retaining the inverse-kinematics solution as a reference.
\subsection{Correspondence Finetuning}
\label{sec:correspondence_refinement}

Minimizing the regularized Chamfer distance, defined in \eqref{eq:chamfer_opt}, brings each SMPL vertex close to the scan surface, but the Chamfer distance maps each template vertex to the nearest scan point independently and does not enforce a one-to-one assignment. Two template vertices may claim the same scan point, and small surface perturbations shift the nearest-neighbor assignment discontinuously. We therefore use $(\bfbeta^*, \bftheta^*)$ as a starting point and replace the Chamfer term with explicit matched-point correspondences, which define a direct point-to-point fitting objective. The details of this correspondence rematching and final parameter fitting are described in the following subsections.
\subsubsection{Correspondence Rematching}
After distance minimization, the fitted SMPL mesh is geometrically close to the scan. We replace the nearest-neighbor associations used by the Chamfer loss with an explicit correspondence map. Let
\begin{equation}
    h = \max_j v_{j,z}-\min_j v_{j,z},
    \qquad
    r = \frac{h}{50},
\end{equation}
where \(h\) is the vertical extent of the fitted SMPL mesh and \(r\) is the initial half-width of the spatial search box.
 
We empirically use \(30{,}000\) scan vertices as the threshold between
dense and sparse 3D scans. We selected this threshold during method development
as the point at which injective matching remained stable without excessive
search failures. Scans above this threshold contain considerably more points
than the \(6{,}890\) vertices of the SMPL template and generally provide enough
distinct scan points to assign one to each template vertex. We therefore
enforce an injective correspondence for dense scans, with each scan point
assigned to at most one template vertex.
 
For each template vertex \(\mathbf{v}_j\), we define a local candidate set using a search box:
\begin{equation}
    \mathcal{N}_{\infty}(\mathbf{v}_j,r_j)
    =
    \left\{
    \mathbf{u}_i:
    \left\|\mathbf{u}_i-\mathbf{v}_j\right\|_{\infty}\leq r_j
    \right\},
\end{equation}
where \(r_j\) is the current half-width of the search box. In the dense mode, we select the closest scan point that has not been assigned to a previously processed template vertex:
\begin{equation}
    C_{\mathrm{dense}}(j)
    =
    \arg\min_{i:\,
    \mathbf{u}_i\in\mathcal{N}_{\infty}(\mathbf{v}_j,r_j),\,
    i\notin\mathcal{A}_{j}}
    \left\|\mathbf{u}_i-\mathbf{v}_j\right\|_2,
\end{equation}
where \(\mathcal{A}_{j}\) contains scan-point indices assigned to template
vertices processed before \(\mathbf{v}_j\). Template vertices are processed in
their fixed SMPL vertex order, making the greedy dense-mode assignment
deterministic. The initial box half-width is \(r_j=r\). If the box contains no
unassigned scan point, we increase its half-width by a factor of \(1.5\). We
perform at most five local searches, using half-widths
\(r,1.5r,\ldots,1.5^4r\). If these searches return no candidate, we select the
globally closest unassigned scan point. This greedy procedure ensures an
injective correspondence. Spatial filtering limits Euclidean distance
calculations to scan points inside a small box around each template vertex.
This strategy substantially reduces the number of computationally expensive
distance evaluations for dense scans, which may contain up to one million
vertices.
 
Scans containing at most \(30{,}000\) vertices may have sparse or nonuniform surface coverage. Enforcing injectivity in this case can force a template vertex to use a distant scan point because a closer point has already been assigned. In the sparse mode, we therefore determine the nearest scan point independently for every template vertex:
\begin{equation}
    C_{\mathrm{sparse}}(j)
    =
    \arg\min_i
    \left\|\mathbf{u}_i-\mathbf{v}_j\right\|_2.
\end{equation}
Previously assigned scan points remain eligible during this search. Multiple template vertices may therefore correspond to the same scan point. Dense mode prioritizes unique assignments, whereas sparse mode prioritizes spatial proximity when scan coverage is limited. BODIESReg selects the mode automatically based on the number of scan vertices. These correspondence-rematching constants were selected empirically during method development and then kept fixed for all datasets and experiments. 
 
When anatomically distinct body parts lie close together, nearest-neighbor associations in the Chamfer loss can connect vertices from different body parts. Examples include touching feet and hands positioned near the torso. If automatic registration fails in such configurations, the user can constrain Chamfer matching using the interactive correspondence selector shown in Figure~\ref{fig:correspondence_selector}.
After pose editing, the user selects a region on the input scan and its corresponding region on the template. The user can repeat this procedure for multiple region pairs. Region boundaries need not be exact because their purpose is to exclude associations with nearby, anatomically unrelated regions. During full-body optimization, we compute the Chamfer loss separately for each selected region pair. Template vertices in a selected region can therefore match only scan vertices in its paired region. We compute an additional Chamfer loss between the remaining unselected scan and template vertices.
\subsubsection{Final Parameter Fitting}
Given the correspondence map $C$, let $\mathbf{p}_j^{\text{scan}} = S[C(j)]$ denote the matched scan point for template vertex $\mathbf{v}_j(\bfbeta, \bftheta)$. We minimize the point-to-point distance~\eqref{eq:final_loss}:
\be\label{eq:final_loss}
\begin{aligned}
&E_{\text{vtx}}(\bfbeta, \bftheta) = \frac{1}{3N}
\sum_{j=1}^{N}\|\mathbf{v}_j-\mathbf{p}_j^{\text{scan}}\|^2, \\
     &(\bfbeta^{**}, \bftheta^{**}) = \arg\min_{\bfbeta,\,\bftheta}\; E_{\text{vtx}}(\bfbeta, \bftheta),
\end{aligned}
\ee
initialized at $(\bfbeta^*, \bftheta^*)$. Once the correspondence map has been estimated, each template vertex has a fixed matched scan point, allowing direct point-to-point fitting instead of nearest-neighbor distance minimization. This final optimization refines the registered mesh toward the scan at the vertex level. After convergence, the correspondence map $C$ is rematched from the updated mesh positions, and parameter optimization is repeated. In principle, this alternation between correspondence rematching and parameter optimization can be iterated arbitrarily, but in our experiments, a single additional pass was sufficient to achieve sub-centimeter mean surface-fit error across the full body surface.

Our registration pipeline is compatible with the parametric body model family, including SMPL~\cite{loper2015smpl}, SMPL+H~\cite{romero2022embodied}, SMPL-X~\cite{pavlakos2019expressive}, and DMPL, which models soft-tissue deformations for SMPL~\cite{loper2015smpl} using a shape space learned from 4D
scans of various subjects in motion. Researchers can therefore select the variant appropriate for their dataset characteristics and downstream use case. The correspondence rematching and parameter optimization stages are agnostic to the specific model variant and allow application across the SMPL family.
\section{Evaluation}
\label{sec:evaluation_error_analysis}
We evaluated BODIESReg on two datasets: CHI3D and MorphoMotion. CHI3D contains synthetic, textureless human meshes in close-interaction poses. We used the CHI3D dataset to test registration under large pose deviations and difficult keypoint detection. MorphoMotion contains real optical body scans acquired in standardized A-pose and sit-pose configurations. We used MorphoMotion to test registration under practical scanning artifacts such as surface noise and missing geometry. We describe both datasets in the next subsection.

We used the same hyperparameters for all datasets. For the pose-initialization stage, we used our adapted VPoser IK framework~\cite{pavlakos2019expressive} (Section~\ref{sec:pose_estimation_initial_mesh}) with the L-BFGS solver, latent dimension $D_z = 32$, and regularization weights $\lambda_z = 0.001$ and $\lambda_\beta = 0.001$. For distance minimization and correspondence refinement, we used the Adam optimizer with a learning rate of $0.005$ and $6000$ iterations without an early-stopping criterion. We performed the CHI3D registrations on a workstation equipped with an NVIDIA RTX A5000 GPU and the MorphoMotion registrations on the DelftBlue supercomputer~\cite{DHPC2024}.
\subsection{Datasets}
\label{sec:datasets}
\subsubsection{CHI3D Dataset}
\label{sec:chi3d_dataset}
Fieraru et al.~\cite{fieraru2020three, fieraru2025reconstructing} introduced the Close-Human-Interaction-3D (CHI3D) dataset as a motion-capture repository for 3D pose and shape reconstruction in close human interaction
scenarios. From their training split (subjects \texttt{s02}, \texttt{s03},
\texttt{s04}), we extracted samples from sequences belonging to eight action categories:
grabbing, handshaking, hitting, holding hands, hugging, kicking, posing, and pushing.
For each qualifying sequence, we took both persons at frame $t = 0$, yielding
$373 \times 2 = 746$ single-person samples. The selection is deterministic and fully
reproducible; the accompanying repository provides a script that reconstructs the exact
746 meshes (10,475~vertices, 20,940~faces) from the SMPL-X
parameters~\cite{pavlakos2019expressive} available on the CHI3D
website~\footnote{\url{https://ci3d.imar.ro/chi3d}}. Because these meshes are generated from parametric ground-truth fits, they are clean, complete, watertight, and noiseless. We used CHI3D to evaluate BODIESReg on complex human poses from close-interaction scenarios. This dataset lets us examine failures driven primarily by pose complexity rather than by scan noise or missing surface geometry. CHI3D also creates a keypoint-detection challenge because its meshes are textureless, whereas MediaPipe~\cite{lugaresi2019mediapipe} is trained on natural human images with texture. This can make automatic keypoint detection less reliable than on textured scans. Since the sex of the subjects was unknown, we used the sex-neutral SMPL-X model for registration, consistent with the CHI3D dataset format. 
\subsubsection{MorphoMotion Dataset}
\label{sec:morphomotion_dataset}
The MorphoMotion dataset~\cite{cueto2026msk} contains 3D body scans from 30 female and 30 male participants measured using the Vitus Bodyscan (VITRONIC Machine Vision GmbH, Wiesbaden, Germany). Participants had a mean age of $51 \pm 18$ years, mean height of $1.74 \pm 0.10$~m, and mean body mass of $73.3 \pm 13.1$~kg. All participants were asked to wear the same model of tight shorts
and sports bra when applicable, and a cap to cover their hair.

Per participant, two 3D scans were acquired in a standardized A-pose: standing
with legs abducted at the hip and supinated arms with shoulder abduction.

One seated 3D scan was acquired per participant: each participant sat on a
standardized stool, slightly abducting the legs to align the knees with the
hips and maintaining an approximate 90-degree knee flexion angle; the arms
were pronated with flexed elbows, and the shoulders rotated until the palms
faced frontally, with adjustments permitted for individual comfort. The dataset also contains whole-body MRI
scans of the same participants acquired in the supine position. 

At the time of this study, data from 50 of the 60 participants were available for
processing, comprising 167 3D surface scans: 114 A-pose scans (57 female, 57 male;
some participants contributed two A-pose scans) and 53 sit-pose scans (25 female, 28 male).
In addition, one MRI-derived skin surface mesh from a single participant was available.
We used sex-specific SMPL+H models (male or female, matched to each participant),
augmented with DMPL to capture soft-tissue deformations. In contrast to CHI3D, MorphoMotion consists of real optical scans acquired in A-pose and sit-pose configurations. We used this dataset to evaluate BODIESReg under practical scanning conditions, including surface noise, missing geometry, and partial scan coverage. A-pose scans occasionally contain amorphous or incorrectly reconstructed hand geometry. Sit-pose scans were acquired with a chair and footrest, which were manually removed before registration, leaving hollow regions around the hips, upper legs, and feet.

The MRI skin surface presents distinct challenges: the participant lay flat during
acquisition, producing a body shape that deviates from the standing-pose distribution on
which SMPL was trained, and the resulting skin surface contains holes and missing regions.
For this case, we relied on the manual fallback tools (the pose editor and correspondence
selector) to initialize registration.
\subsection{Evaluation Metrics}
\label{sec:eval_metrics}
\subsubsection{Failure Criterion}
\label{sec:joint_failure_criterion}
When the ground-truth joint positions of the 3D scans are available, we assess registration failure from the discrepancy between the ground-truth and registered skeletons. We first remove residual global translation by aligning their pelvis positions. Let $\mathcal{J}$ denote the set of 22 evaluated body joints, including the pelvis and extending through the wrists. For each $j\in\mathcal{J}$, let $Q_j$ and $\hat{Q}_j$ denote the registered and ground-truth joint positions, respectively. We define the maximum joint error as
\begin{equation}
 e_{\max}=\max_{j\in\mathcal{J}}\bigl\|Q_j-\hat{Q}_j\bigr\|_2.
\label{eq:max_joint_error}
\end{equation}
For each dataset with ground-truth joints, we sort the registrations by $e_{\max}$ in ascending order and identify the transition region in the sorted error curve. Starting from the low-error side of this region, we visually inspect the registrations and define the failure boundary at the first anatomically incorrect registration. Registrations at or above this boundary are classified as failed. This procedure avoids imposing an ad hoc numerical threshold on joint error for failure that may not transfer between datasets.

The CHI3D dataset meshes are generated from known SMPL-X parameters. We recover their ground-truth posed joint positions through a forward pass of the SMPL-X model and obtain the registered joint positions through a forward pass of the parameters of the registered model. We report the resulting CHI3D failure boundary in Section~\ref{sec:results_failure}.

When ground-truth joint positions are unavailable, $e_{\max}$ cannot be computed. MorphoMotion scans and the MRI-derived skin surface have no ground-truth joint positions, so we assess their anatomical correctness by visual inspection. The joint-error criterion is therefore used only to evaluate datasets with ground-truth skeletons. 
\subsubsection{Surface-Fit Error Metric} 
\label{sec:accuracy_metric}
For registrations classified as anatomically correct by the failure assessment,
we report per-segment surface-fit error. In Stage~3 (Correspondence Finetuning,
Section~\ref{sec:correspondence_refinement}), we establish a correspondence map
$C(j)$ that assigns each optimized template vertex $\mathbf{v}_j$ to a point
$\mathbf{p}_j^{\text{scan}}$ on the target scan surface. For dense scans, this
map is injective, so each scan point is assigned to at most one template vertex.
For sparse scans, the map may be many-to-one, so multiple template vertices may
be assigned to the same scan point. We define the per-vertex surface-fit error as
$\|\mathbf{v}_j(\bfbeta^{**},\bftheta^{**}) - \mathbf{p}_j^{\text{scan}}\|$
and report the mean $\pm$ standard deviation, median, 95th percentile, and maximum
error per segment. Because the scan-surface points are selected from correspondences
established during fitting, this measure is an in-sample fitting residual and does
not independently validate anatomical correspondence. Anatomical correctness is
assessed separately using the joint-based or visual failure assessment described
in Section~\ref{sec:joint_failure_criterion}. For each segment, surface-fit errors
are pooled across all registered scans; for MorphoMotion, pooling is performed
separately for each sex and pose group.
\subsection{Results} 
\label{sec:results}
\subsubsection{Failure Analysis} 
\label{sec:results_failure}
\begin{figure}
    \centering
   \begin{overpic}[width=0.5\linewidth,  unit=2bp,tics=7 ] {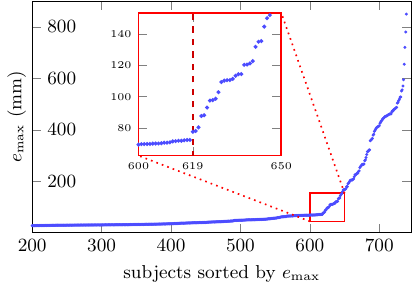}
    \end{overpic}
\caption{ Maximum joint error $e_{\max}$ for all 746 CHI3D registrations,
sorted in ascending order. The inset shows the transition region
(ranks 600--650) where $e_{\max}$ rises sharply; registrations above
rank 618 are classified as failures and excluded from accuracy
evaluation.} 
        \label{fig:max_joint_error}
\end{figure}
\begin{figure}
    \centering
   \begin{overpic}[width=0.5\linewidth,  unit=2bp,tics=7  ] {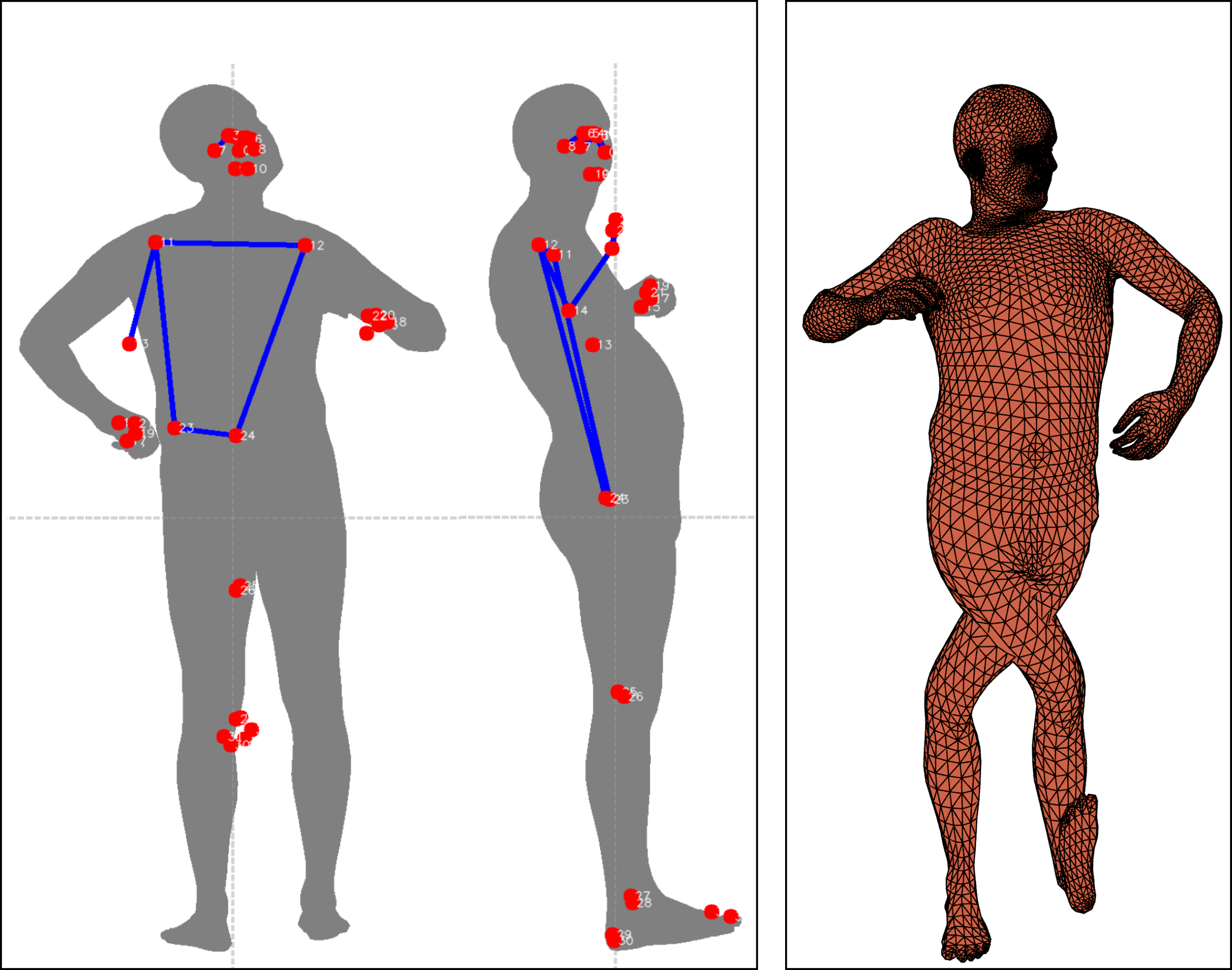}
       \put(12,75){$\scriptstyle{\text{keypoint detection failure}}$}
  \put(68,75){$\scriptstyle{\text{registration failure}}$}
    \end{overpic}
\caption{Illustration of the cascading failure in the CHI3D dataset. Left: front and side
views of a standing subject; keypoint detection (red dots) fails to
localize lower-limb joints and elbows accurately. Right: the
resulting registration failure, where the erroneous pose
initialization drives the mesh into a running configuration with
raised knee and twisted ankles, severely misaligning it with the
standing-pose scan.}
        \label{fig:failed_CHI3D}
\end{figure}
\begin{figure*}
    \centering  
   \begin{overpic}[width=0.86\linewidth,  unit=2bp,tics=7 ] {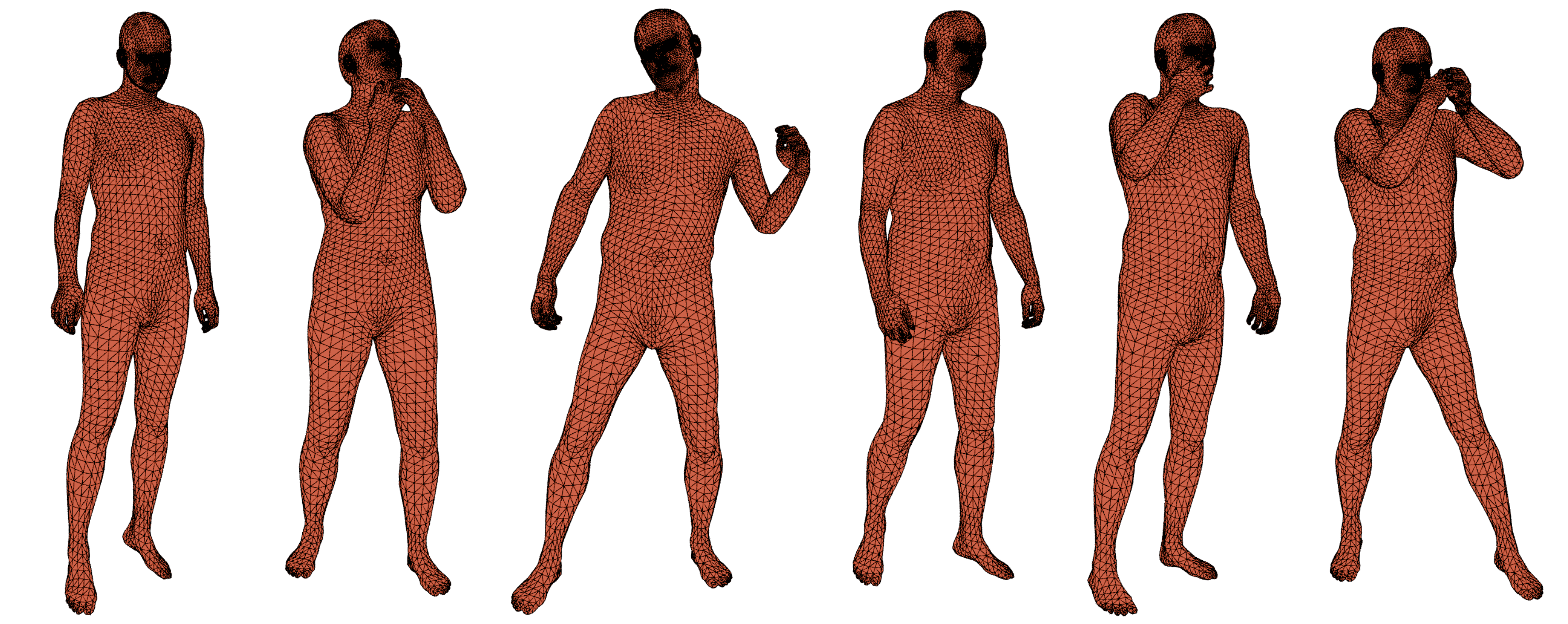}
     \end{overpic}
\caption{Representative examples of CHI3D registrations using the automatic pipeline.}
        \label{fig:CHI3DAutomatic}
\end{figure*}

In Figure~\ref{fig:max_joint_error}, we show $e_{\max}$, defined in~\eqref{eq:max_joint_error}, for all 746 CHI3D registrations sorted in ascending order. The curve remains low for most successful registrations and then increases steeply near rank 619, indicating a transition between correct and failed registrations at rank 619. Failures
correspond to cases where the optimizer converged to an incorrect pose,
typically with swapped or reflected limbs.  Visual inspection confirmed
that all 618 registrations below the transition are correctly
registered. We also inspected the entire transition region shown in the inset of Figure~\ref{fig:max_joint_error} to verify the failures. In Figure~\ref{fig:failed_CHI3D}, we show a representative case in which pose-initialization failure led to registration failure. Representative examples of successful CHI3D registrations are shown in Figure~\ref{fig:CHI3DAutomatic}.

No registration in the MorphoMotion dataset failed visual inspection. This result is consistent with the availability of texture in the scans, which supported keypoint detection and pose estimation.
%

%
%
\subsubsection{Surface-Fit Error} 
\label{sec:results_accuracy}
For the CHI3D dataset, per-segment surface-fit errors across the 618 anatomically correct registrations are reported in Table~\ref{tab:chi3d_segment_errors}. Mean errors were below 9~mm for all segments and ranged from 2.0~mm (Right Foot) to 8.6~mm (Hips). Leg, foot, and toe-base segments had mean errors of 2.0--2.4~mm, head and neck segments had mean errors of 3.0--3.2~mm, and hips and upper-leg segments had mean errors of 6.9--8.6~mm. The Right Foot standard deviation nearly equals its mean ($2.03\pm2.02$~mm), indicating a subset of registrations with large foot surface-fit errors. The 95th-percentile errors remain below 19~mm for all segments. The maximum
error across segments remains below 100~mm.
 \begin{figure*}
    \centering  
   \begin{overpic}[width=1\linewidth,  unit=2bp,tics=7 ] {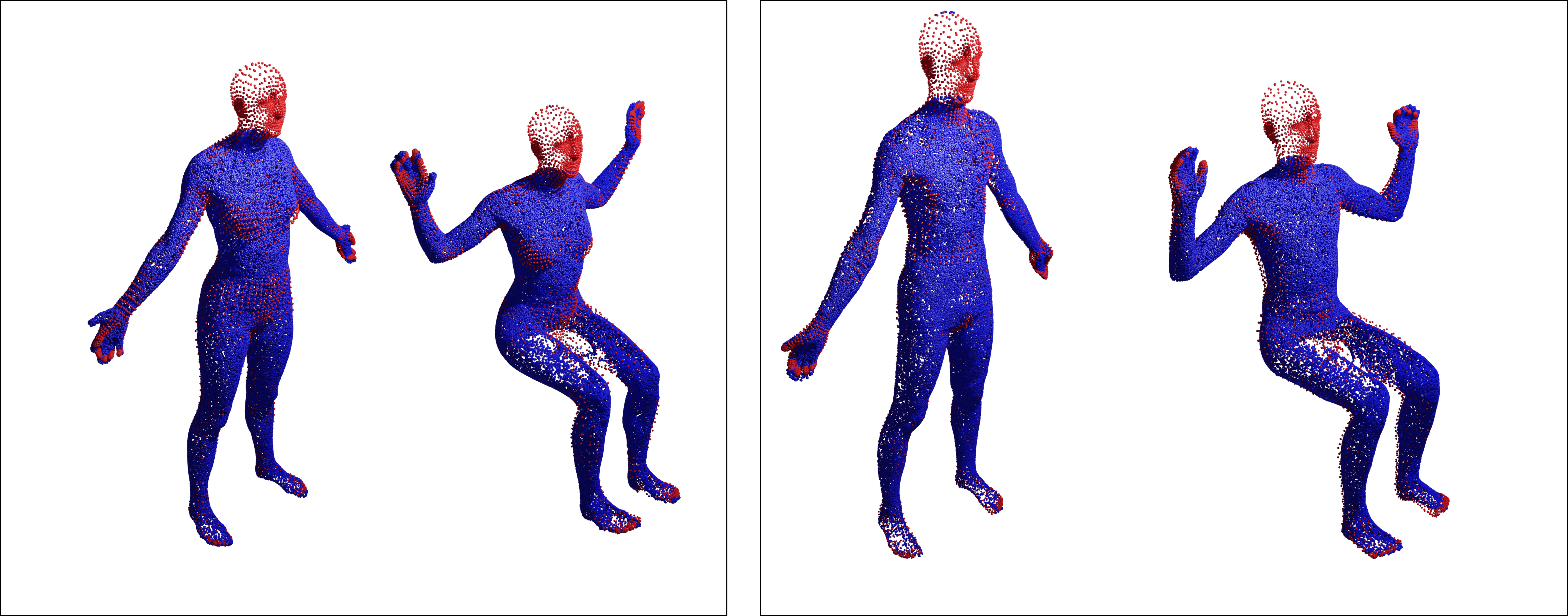}

  \put(19,35.7){  Female}
   \put(12,0.95){{  A-pose}}
   \put(32,0.95){{  sit-pose}}
    \put(72,35.7){  Male}
   \put(57,0.95){{    A-pose}}
    \put(82,0.95){{   sit-pose}}
     
   \end{overpic}
\caption{Representative scans from the MorphoMotion dataset and their registered point clouds (shown in red). From left to right: female A-pose, female sit-pose, male A-pose, and male sit-pose. Heads are hidden in the displayed 3D scans for anonymity but were included during registration. In the sit-pose scans, some parts of the hip, upper legs, and feet are missing due to occlusion by the chair and footrest, which were manually removed from the scans.}
        \label{fig:morphomotion_registration}
\end{figure*}
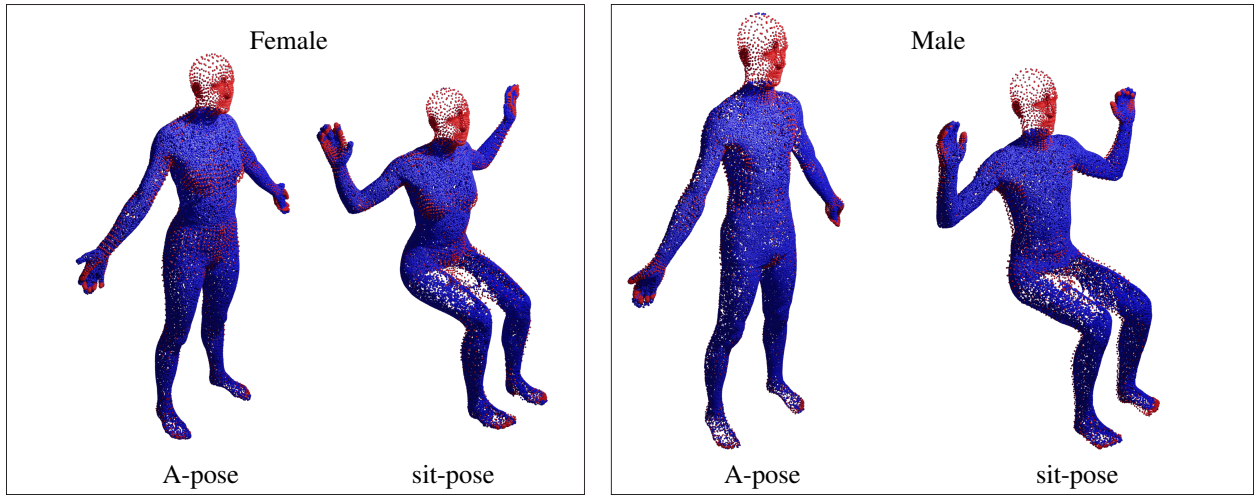
\begin{table}[!ht]
\centering
\caption{CHI3D surface-fit error by anatomical segment: mean~$\pm$~standard deviation, median, 95th percentile, and maximum template-vertex-to-scan-surface distance (mm) across 618 anatomically correct automatic registrations (128 pose failures excluded; see Section~\ref{sec:joint_failure_criterion}). Segments are ordered anatomically from top to bottom.}
\label{tab:chi3d_segment_errors}
\vspace{0.2in}
\renewcommand{\arraystretch}{1.1}
\footnotesize
\begin{tabular*}{0.72\textwidth}{@{}l@{\extracolsep{\fill}}cccc@{}}
\toprule
\textbf{Segment} & \textbf{Mean$\pm$Std} & \textbf{Med} & \textbf{P95} & \textbf{Max} \\
\midrule
Head              & $2.99\pm2.61$ & 2.36 & 7.39 & 47.61 \\
Neck              & $3.15\pm1.95$ & 2.74 & 6.62 & 27.22 \\
\midrule
Spine 1           & $5.59\pm3.79$ & 4.77 & 12.15 & 96.02 \\
Spine 2           & $4.94\pm3.50$ & 4.09 & 11.20 & 79.09 \\
\midrule
Spine             & $7.05\pm4.57$ & 6.55 & 14.47 & 96.02 \\
Hips              & $8.60\pm5.82$ & 7.74 & 18.75 & 65.93 \\
\midrule
Right Shoulder    & $6.24\pm4.33$ & 5.36 & 14.46 & 58.22 \\
Left Shoulder     & $6.19\pm3.78$ & 5.69 & 12.73 & 44.05 \\
\midrule
Right Arm         & $5.25\pm3.94$ & 4.48 & 11.81 & 60.26 \\
Left Arm          & $5.83\pm4.18$ & 5.12 & 12.53 & 74.19 \\
\midrule
Right Forearm     & $4.49\pm3.89$ & 3.81 & 9.33 & 82.57 \\
Left Forearm      & $5.10\pm5.19$ & 4.06 & 11.49 & 81.92 \\
\midrule
Right Hand        & $6.44\pm5.05$ & 5.20 & 15.82 & 63.39 \\
Left Hand         & $6.91\pm5.22$ & 5.59 & 16.72 & 67.40 \\
\midrule
Right Hand Index  & $6.45\pm5.22$ & 4.73 & 17.50 & 43.48 \\
Left Hand Index   & $6.44\pm5.06$ & 4.82 & 17.06 & 43.67 \\
\midrule
Right Upper Leg   & $7.48\pm5.92$ & 5.94 & 18.47 & 63.18 \\
Left Upper Leg    & $6.90\pm5.30$ & 5.59 & 16.58 & 57.04 \\
\midrule
Right Leg         & $2.38\pm2.07$ & 1.85 & 6.06 & 33.20 \\
Left Leg          & $2.35\pm1.96$ & 1.85 & 5.99 & 28.71 \\
\midrule
Right Foot        & $2.03\pm2.02$ & 1.61 & 4.23 & 31.88 \\
Left Foot         & $2.06\pm1.94$ & 1.65 & 4.50 & 29.14 \\
\midrule
Right Toe Base    & $2.09\pm1.24$ & 1.97 & 3.84 & 15.71 \\
Left Toe Base     & $2.12\pm1.25$ & 2.03 & 3.97 & 16.37 \\
\bottomrule
\end{tabular*}
\end{table}
 For the MorphoMotion dataset, per-segment errors across 167 scans are reported in
Table~\ref{tab:combined_segment_accuracy}, separated by sex and pose. In A-pose, all 24 segments had sub-centimeter mean errors across both sexes.
Leg and forearm segment mean errors were 4.1--5.6~mm, and foot segment mean errors were 8.0--8.7~mm.
For the Right Foot, the standard deviations were $7.16$~mm for females and $7.54$~mm for males, indicating occasional large outliers from noise near the foot-ground contact region. 
The 95th-percentile errors remain below 20~mm for all trunk and arm segments;
foot P95 values reach 24--26~mm. Maximum errors stay below 83~mm across all A-pose segments. 

In sit-pose scans, the chair and foot wedge were manually removed before registration. This cleaning step also removed scan data in contact regions around the hips, upper legs, feet, and toe bases. We therefore do not report surface-fit errors for these segments in sit-pose scans (shown as~\mbox{NA}; see Section~\ref{sec:morphomotion_dataset}). For the remaining sit-pose segments, mean errors range from 2.4 to 7.8~mm across sexes. Representative examples of MorphoMotion registrations are shown in Figure~\ref{fig:morphomotion_registration}.
\begin{sidewaystable*}
\centering
\caption{MorphoMotion surface-fit error by anatomical segment across pose and sex groups. Mean $\pm$ standard deviation, median, 95th percentile (P95), and maximum template-vertex-to-scan-surface distances are reported in millimeters (mm) for 167 scans: 57 female A-pose, 57 male A-pose, 25 female sit-pose, and 28 male sit-pose. In sit-pose scans, entries marked~\mbox{NA} correspond to segments where scan data were removed during manual cleaning of the chair and foot wedge before registration. Across all reported segments, poses, and sex groups, the mean surface-fit error is below 10~mm.}
\label{tab:combined_segment_accuracy}
\vspace{0.2in}
\scriptsize
\renewcommand{\arraystretch}{1.5}
\setlength{\tabcolsep}{5pt}
\begin{tabular}{@{}l rrrr rrrr rrrr rrrr@{}}
\toprule
\textbf{Segment}
  & \multicolumn{4}{c}{\textbf{Female A-pose (57)}}
  & \multicolumn{4}{c}{\textbf{Male A-pose (57)}}
  & \multicolumn{4}{c}{\textbf{Female Sit-pose (25)}}
  & \multicolumn{4}{c}{\textbf{Male Sit-pose (28)}} \\
\cmidrule(lr){2-5}\cmidrule(lr){6-9}\cmidrule(lr){10-13}\cmidrule(lr){14-17}
 & \textbf{Mean$\pm$Std} & \textbf{Med} & \textbf{P95} & \textbf{Max}
 & \textbf{Mean$\pm$Std} & \textbf{Med} & \textbf{P95} & \textbf{Max}
 & \textbf{Mean$\pm$Std} & \textbf{Med} & \textbf{P95} & \textbf{Max}
 & \textbf{Mean$\pm$Std} & \textbf{Med} & \textbf{P95} & \textbf{Max} \\
\midrule
Head              & $4.22\pm3.08$ & 3.38 & 10.56 & 46.61 & $3.98\pm2.53$ & 3.45 & 8.68 & 34.02 & $3.86\pm2.95$ & 3.06 & 9.55 & 36.77 & $3.81\pm2.37$ & 3.35 & 8.34 & 31.56 \\
Neck              & $4.75\pm3.04$ & 4.04 & 10.80 & 20.33 & $4.67\pm2.92$ & 4.11 & 10.25 & 23.23 & $4.25\pm2.91$ & 3.54 & 10.23 & 21.40 & $4.63\pm2.89$ & 3.95 & 10.41 & 20.42 \\
\midrule
Spine 1           & $5.02\pm3.49$ & 4.15 & 12.08 & 28.89 & $5.23\pm3.57$ & 4.41 & 12.09 & 25.66 & $5.09\pm3.80$ & 4.08 & 12.83 & 23.89 & $5.29\pm3.49$ & 4.47 & 12.01 & 26.36 \\
Spine 2           & $5.86\pm5.53$ & 4.31 & 15.85 & 61.29 & $6.09\pm5.78$ & 4.68 & 15.76 & 80.68 & $4.89\pm3.68$ & 3.96 & 12.05 & 36.41 & $5.19\pm3.60$ & 4.26 & 12.17 & 41.38 \\
\midrule
Spine             & $5.58\pm3.75$ & 4.70 & 12.93 & 24.35 & $5.26\pm3.62$ & 4.47 & 12.03 & 29.86 & $5.61\pm4.00$ & 4.63 & 13.33 & 24.06 & $6.40\pm4.97$ & 4.95 & 16.32 & 38.62 \\
Hips              & $6.02\pm4.52$ & 4.95 & 14.26 & 42.82 & $6.36\pm4.96$ & 5.21 & 15.19 & 61.99 & \multicolumn{4}{c}{NA} & \multicolumn{4}{c}{NA} \\
\midrule
Right Shoulder    & $4.83\pm3.31$ & 4.03 & 11.28 & 34.14 & $4.95\pm3.49$ & 4.11 & 11.66 & 34.28 & $4.33\pm3.10$ & 3.51 & 10.81 & 20.56 & $4.90\pm3.25$ & 4.06 & 11.59 & 22.81 \\
Left Shoulder     & $4.78\pm3.49$ & 3.96 & 11.15 & 43.31 & $5.78\pm4.60$ & 4.68 & 14.41 & 62.44 & $4.46\pm2.94$ & 3.83 & 10.48 & 26.46 & $5.00\pm3.34$ & 4.18 & 11.58 & 29.66 \\
\midrule
Right Arm         & $5.67\pm4.89$ & 4.48 & 13.52 & 50.78 & $5.43\pm5.06$ & 4.16 & 13.67 & 53.49 & $5.36\pm4.24$ & 4.17 & 13.34 & 43.32 & $5.54\pm4.12$ & 4.55 & 12.18 & 48.85 \\
Left Arm          & $5.87\pm5.58$ & 4.43 & 14.90 & 55.96 & $6.69\pm7.11$ & 4.72 & 19.13 & 82.31 & $5.11\pm4.10$ & 4.11 & 12.18 & 38.96 & $5.67\pm4.46$ & 4.51 & 14.04 & 45.78 \\
\midrule
Right Forearm     & $5.58\pm3.87$ & 4.56 & 13.26 & 25.32 & $4.31\pm2.96$ & 3.70 & 9.91 & 25.62 & $3.95\pm3.36$ & 3.14 & 9.25 & 43.32 & $4.27\pm3.36$ & 3.52 & 9.65 & 45.22 \\
Left Forearm      & $5.31\pm4.13$ & 4.17 & 13.23 & 42.73 & $5.38\pm4.58$ & 4.05 & 14.27 & 40.09 & $4.23\pm3.55$ & 3.36 & 10.65 & 38.96 & $4.40\pm3.62$ & 3.58 & 10.08 & 45.06 \\
\midrule
Right Hand        & $6.10\pm4.16$ & 5.01 & 14.55 & 27.04 & $6.07\pm4.37$ & 4.93 & 14.87 & 33.69 & $4.23\pm3.03$ & 3.56 & 9.51 & 31.43 & $4.77\pm3.19$ & 4.08 & 10.51 & 33.36 \\
Left Hand         & $5.77\pm4.12$ & 4.74 & 13.69 & 39.92 & $6.71\pm4.92$ & 5.47 & 16.55 & 34.05 & $4.34\pm2.97$ & 3.67 & 9.88 & 21.77 & $3.95\pm2.76$ & 3.35 & 8.85 & 29.96 \\
\midrule
Right Hand Index  & $4.93\pm3.33$ & 4.10 & 11.59 & 24.53 & $4.93\pm3.16$ & 4.31 & 10.95 & 31.72 & $3.71\pm2.24$ & 3.24 & 8.06 & 14.06 & $4.00\pm2.50$ & 3.56 & 8.54 & 26.79 \\
Left Hand Index   & $5.05\pm3.72$ & 4.13 & 11.81 & 40.97 & $5.60\pm3.90$ & 4.71 & 13.10 & 32.35 & $3.72\pm2.29$ & 3.28 & 7.95 & 20.31 & $3.86\pm2.38$ & 3.43 & 8.11 & 23.91 \\
\midrule
Right Upper Leg   & $5.31\pm4.37$ & 4.20 & 12.79 & 42.28 & $5.37\pm4.78$ & 4.35 & 12.49 & 61.56 & \multicolumn{4}{c}{NA} & \multicolumn{4}{c}{NA} \\
Left Upper Leg    & $5.18\pm4.27$ & 4.07 & 12.44 & 42.13 & $5.47\pm4.80$ & 4.45 & 12.98 & 61.75 & \multicolumn{4}{c}{NA} & \multicolumn{4}{c}{NA} \\
\midrule
Right Leg         & $4.24\pm2.73$ & 3.60 & 9.73 & 20.81 & $4.31\pm2.57$ & 3.89 & 9.10 & 20.37 & $4.79\pm3.54$ & 3.83 & 11.83 & 25.10 & $4.66\pm3.07$ & 3.99 & 10.78 & 25.34 \\
Left Leg          & $4.10\pm2.69$ & 3.53 & 9.31 & 21.53 & $4.30\pm2.70$ & 3.78 & 9.54 & 21.36 & $5.25\pm3.54$ & 4.40 & 12.19 & 27.93 & $5.30\pm3.46$ & 4.50 & 11.88 & 26.26 \\
\midrule
Right Foot        & $8.35\pm7.16$ & 5.68 & 24.30 & 42.30 & $8.68\pm7.54$ & 5.82 & 25.72 & 42.33 & \multicolumn{4}{c}{NA} & \multicolumn{4}{c}{NA} \\
Left Foot         & $8.03\pm7.08$ & 5.28 & 23.89 & 38.33 & $8.56\pm7.52$ & 5.73 & 25.52 & 45.24 & \multicolumn{4}{c}{NA} & \multicolumn{4}{c}{NA} \\
\midrule
Right Toe Base    & $6.32\pm4.96$ & 4.91 & 15.83 & 34.25 & $6.54\pm5.05$ & 5.18 & 16.18 & 37.44 & \multicolumn{4}{c}{NA} & \multicolumn{4}{c}{NA} \\
Left Toe Base     & $6.24\pm4.97$ & 4.78 & 15.87 & 32.94 & $6.35\pm4.81$ & 5.22 & 15.53 & 33.52 & \multicolumn{4}{c}{NA} & \multicolumn{4}{c}{NA} \\
\bottomrule
\end{tabular}
\vspace{5pt}
\end{sidewaystable*}
We report per-vertex surface-fit errors from a single representative
MRI-derived skin-surface registration in Table~\ref{tab:mri_segment_errors}. 
Unlike the CHI3D and MorphoMotion summaries, these values describe one
registration rather than performance across multiple scans. Mean surface-fit
errors are higher than those for CHI3D and MorphoMotion
(Tables~\ref{tab:chi3d_segment_errors} and
\ref{tab:combined_segment_accuracy}). For this case, 14 of 24 segments have
mean errors below 10~mm. For the remaining 10 segments, including the trunk,
arms, forearms, left upper leg, right foot, and right toe base, mean errors
range from 10.7 to 15.4~mm. In Figure~\ref{fig:mriSkinRegistration}, we show
the MRI-derived skin surface and the registered mesh superimposed. We attribute the higher errors in these 10 segments mainly to two aspects of the scan. First, some regions of
the MRI-derived skin surface are missing, especially around the forearms,
hands, and posterior body surface. In these regions, template vertices are
assigned to the nearest available point on an adjacent surface, which may not
be the anatomically corresponding surface point. Second, contact with the firm
support during supine MRI acquisition flattens the posterior soft tissue,
which can fall outside the SMPL shape space. 

For datasets where scans share a similar pose, our pipeline supports
fully automatic batch processing by using a single registered mesh as
the initial template, removing the need for manual initialization on
subsequent scans. 
\subsection{Computational Performance}
 We measured registration time for both datasets used in this study. The
CHI3D dataset, described in Section~\ref{sec:chi3d_dataset}, contains 746
scans with a mean file size of 740~KB. The registrations were performed on a
workstation with an NVIDIA RTX A5000 GPU, and the mean runtime was
210~seconds per registration. The MorphoMotion dataset, described in
Section~\ref{sec:morphomotion_dataset}, contains 167 scans with a mean file
size of 250~MB. The registrations were performed on the DelftBlue
supercomputer~\cite{DHPC2024}, and the mean runtime was 500~seconds per
registration. 
 
\subsection{Acceleration Settings}
For machines with limited memory or GPU, BODIESReg provides
two acceleration settings: memory-efficient Chamfer evaluation and hybrid
optimization. The first setting combines blockwise distance evaluation with an
optional one-directional Chamfer distance. Users can enable the two settings
independently or together; the default CPU configuration uses both.

\textbf{(a) Memory-efficient Chamfer evaluation.}
Pairwise-distance evaluation within the bidirectional Chamfer objective in \eqref{eq:chamfer_opt} is the main memory cost during optimization. For $M$ template vertices and $N$ scan vertices, direct evaluation stores an $M\times N$ distance matrix and requires $\mathcal{O}(MN)$ temporary memory.

To reduce memory usage, users can divide both point sets into blocks of at most $C$ points each. We compute one $C\times C$ distance block at a time and retain only the minimum distances required to calculate the loss. Therefore, the largest temporary distance matrix contains at most $C^2$ values. Chunking changes memory use but does not change the Chamfer objective or the total number of point-pair comparisons. The default CPU configuration uses $C=512$.

Users can further reduce computation by selecting one-directional Chamfer distance:
\be
E_{\mathrm{dir}}(\bfbeta,\bftheta)
=
\frac{1}{|\mathcal{V}|}
\sum_{\bfv\in\mathcal{V}}
\min_{\bfu\in S}
\|\bfv-\bfu\|.
\ee
One-directional Chamfer distance retains the template-to-scan term and removes the scan-to-template pass, requiring approximately half as many distance-block evaluations. Users can combine chunking with either symmetric or one-directional Chamfer distance.

\textbf{(b) Hybrid optimization.}
Users can divide the optimization between Adam and L-BFGS. We use Adam during the initial iterations because it is less sensitive to imperfect initialization and noisy gradients. We then use the Adam output to initialize L-BFGS, which refines the solution near a local minimum. Users specify the total number of iterations and the number assigned to Adam; the remaining iterations are assigned to L-BFGS. Assigning all iterations to Adam retains Adam-only optimization.

We randomly selected 10 scans from CHI3D and 10 scans from MorphoMotion to evaluate the default CPU configuration. We tested BODIESReg on an Intel CPU with 16~GB of RAM and no dedicated GPU. We used one-directional Chamfer distance with chunk size $C=512$ and 2000 optimization iterations, comprising 100 Adam iterations followed by 1900 L-BFGS iterations. All registrations achieved mean surface-fit errors below 10~mm. The average runtime was $72$~seconds per CHI3D scan and $240$~seconds per MorphoMotion scan.

The acceleration settings described above involve trade-offs. One-directional Chamfer distance imposes a weaker coverage constraint because scan points without a nearby template vertex do not contribute directly to the loss. L-BFGS can also be sensitive to noisy data or poor initialization, although the initial Adam iterations reduce this sensitivity. A study of these trade-offs is beyond the scope of this work. We recommend the default accelerated configuration as a starting point. If registration is unstable, users can increase the number of Adam iterations or select the symmetric Chamfer distance.
\begin{table}[!ht]
\centering
\caption{MRI skin surface-fit error by anatomical segment: mean~$\pm$~standard deviation of template-vertex-to-scan-surface distances (mm) from a single registration. Segments are ordered anatomically from top to bottom.}
\label{tab:mri_segment_errors}
\vspace{0.2in}
\renewcommand{\arraystretch}{1.1}
\small
\begin{tabular*}{0.52\textwidth}{@{}l@{\extracolsep{\fill}}c@{}}
\toprule
\textbf{Segment} & \textbf{Mean$\pm$Std (mm)} \\
\midrule
Head              & $9.4\pm4.4$ \\
Neck              & $8.3\pm3.8$ \\
\midrule
Spine 1           & $8.6\pm3.3$ \\
Spine 2           & $10.7\pm6.6$ \\
Spine             & $11.8\pm3.6$ \\
Hips              & $11.7\pm5.0$ \\
\midrule
Right Shoulder    & $8.1\pm3.6$ \\
Left Shoulder     & $9.5\pm4.6$ \\
\midrule
Right Arm         & $11.1\pm5.7$ \\
Left Arm          & $14.3\pm9.1$ \\
\midrule
Right Forearm     & $14.0\pm9.1$ \\
Left Forearm      & $15.4\pm10.9$ \\
\midrule
Right Hand        & $8.3\pm3.6$ \\
Left Hand         & $8.8\pm4.2$ \\
\midrule
Right Hand Index  & $6.4\pm2.5$ \\
Left Hand Index   & $6.6\pm2.5$ \\
\midrule
Right Upper Leg   & $9.7\pm4.3$ \\
Left Upper Leg    & $12.0\pm5.2$ \\
\midrule
Right Leg         & $8.1\pm3.1$ \\
Left Leg          & $7.9\pm2.6$ \\
\midrule
Right Foot        & $11.9\pm5.6$ \\
Left Foot         & $7.6\pm3.1$ \\
\midrule
Right Toe Base    & $11.1\pm7.1$ \\
Left Toe Base     & $7.5\pm3.2$ \\
\bottomrule
\end{tabular*}
\end{table}
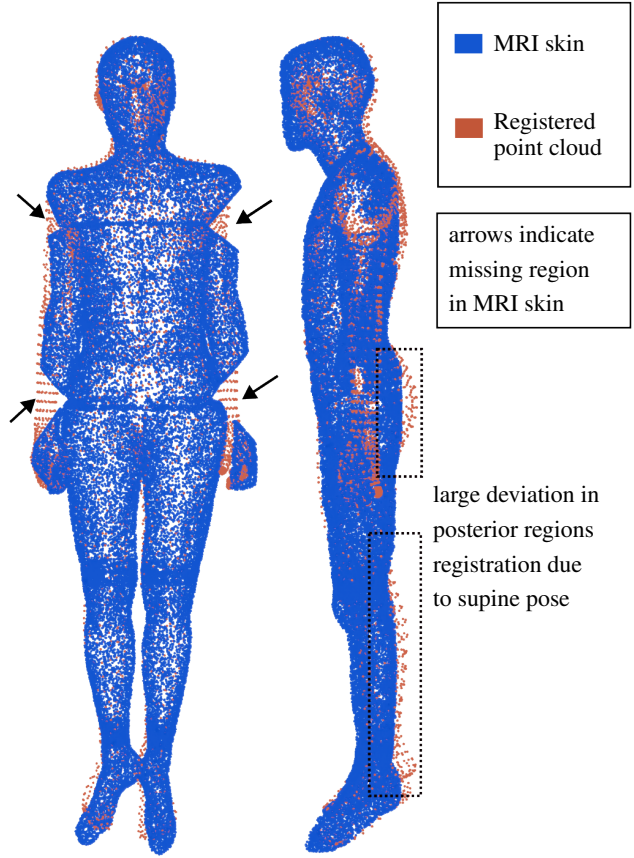
\begin{figure}[ht!]
    \centering  
   \begin{overpic}[width=0.5\linewidth,  unit=2bp,tics=4 ] {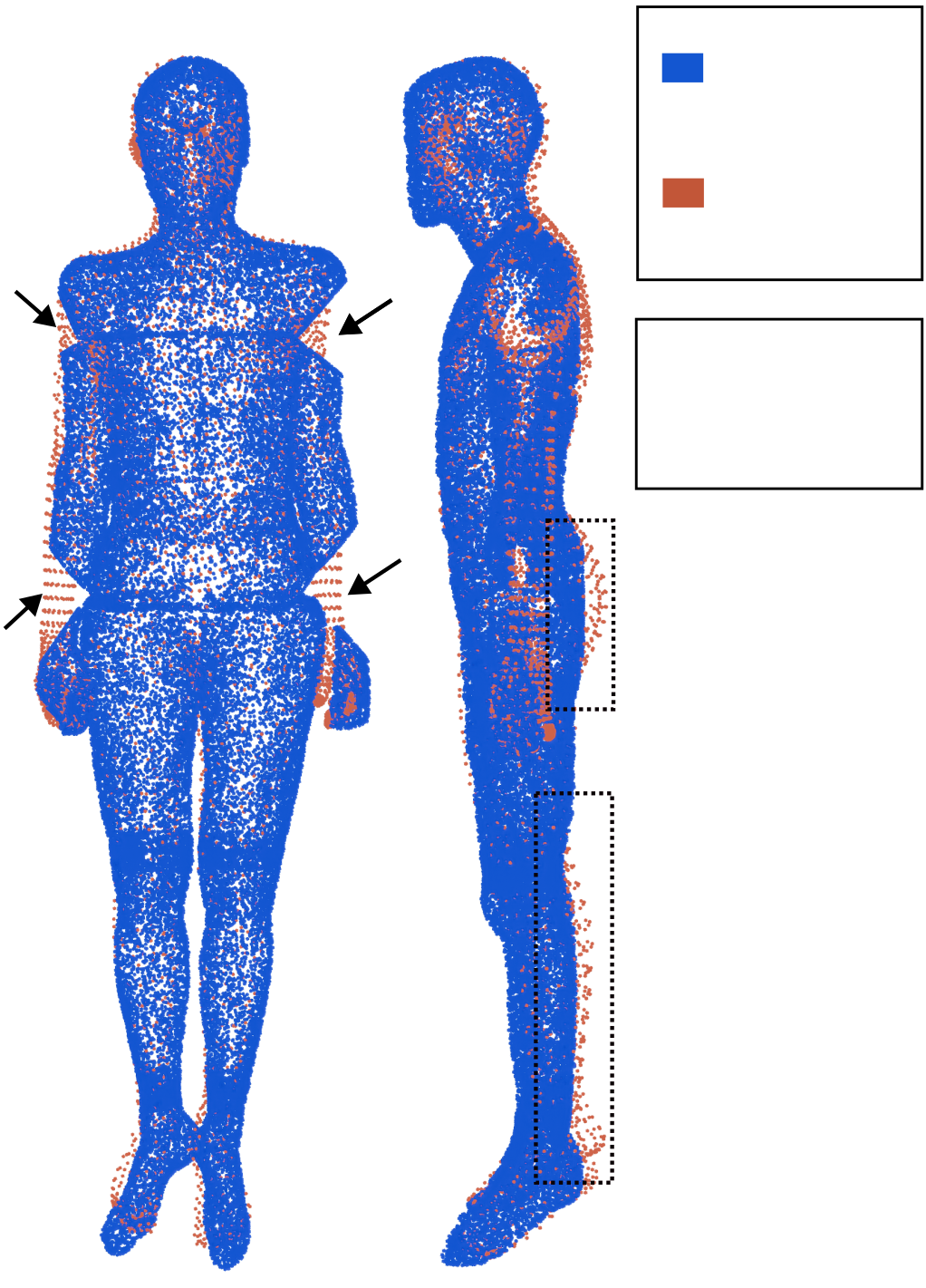}
    \put(56,94){\small MRI skin}
  \put(56,85){\small Registered}
    \put(56,82.0){\small point cloud}
     \put(50.8,72){\small arrows indicate}
        \put(50.8,68){\small  missing region}
 \put(50.8,64){\small in MRI skin}

  \put(49,42){\small large deviation in}
        \put(49,38){\small posterior regions}
          \put(49,34){\small registration due}
        \put(49,30){\small to supine pose}
  \end{overpic}
\caption{Representative MRI skin surface registration. Front and side
views show the input skin surface (blue) and registered point cloud
(orange) superimposed. Arrows indicate missing scan regions where
template vertices are assigned to the nearest adjacent surface, making
per-segment errors there unreliable. The dashed boxes in the side view show large posterior deviations associated with contact-induced soft-tissue flattening against the support
surface during supine MRI acquisition.} 
        \label{fig:mriSkinRegistration}
\end{figure}
  \section{Discussion and Conclusion}
\label{sec:discussion}
We presented BODIESReg, an open-source pipeline for registering a
parametric body model to 3D surface scans. Standard SMPL fitting commonly
starts from a T-pose template. When the scan is far from this pose, the
optimizer can converge to a local minimum and produce anatomically incorrect
meshes. We addressed this issue by detecting anatomical keypoints on the scan,
solving an inverse kinematics problem to recover pose parameters, and
deforming the template to match the estimated pose before surface-distance
minimization.

Pose-aligned initialization, spatially filtered correspondence
rematching, and fixed-correspondence refinement produced mean per-segment
surface-fit errors below 10~mm across pose-diverse datasets. In the ablation
study reported in Appendix~\ref{sec:ablation}, removing pose-aligned
initialization reduced registration success from $82.9\%$ to $9.7\%$ for CHI3D
and from $100\%$ to $71.9\%$ for MorphoMotion. These results indicate that pose-aligned initialization is important when the
target pose differs from the default T-pose. The analysis isolates the
initialization stage; correspondence rematching and fixed-correspondence
refinement remained part of the full pipeline and were not ablated separately. For cases where automatic keypoint detection
fails, BODIESReg includes a pose editor for manual correction of the
initial template. BODIESReg also includes a post-processing tool that automatically
extracts segmental volumes, cross-sectional measurements, and segment lengths from registered meshes. We provide implementation details for this tool in Appendix~\ref{sec:appendix_anthropometrics}.
%
%
\subsection{Limitations}
\label{sec:discussion_limitations}
Our pipeline inherits limitations from its design and from the underlying SMPL model. We organize these limitations by their effects on anatomical correctness and surface fit.

\subsubsection{Pose-Initialization Sensitivity}
Anatomical correctness depends fundamentally on the quality of the
pose-aligned initial template. The pose-initialization stage uses
MediaPipe for keypoint detection, and pose quality depends on three
factors: scan surface quality (noise, holes, missing regions), pose
complexity (extreme articulations, self-occlusions), and MediaPipe
detection accuracy. MediaPipe was trained on natural colored images,
so scans without color or texture information produce unreliable
keypoint detections and may require manual correction via the
pose editor.

Keypoint detection errors are most pronounced for textureless or
uncolored inputs; colored surface meshes and RGB scans are less affected. If
automatic detection fails, users can correct the initial pose manually with the
pose editor before registration proceeds. 

\subsubsection{Accuracy-Robustness Trade-off}
The distance minimization function we chose is weakly regularized with
respect to pose and shape parameters and does not include a statistical shape
or pose prior. This weak regularization allows the optimizer to follow the scan
surface more closely, which can improve surface-fit accuracy when the
pose-aligned initialization is correct. The trade-off is reduced protection
against poor initialization.

If the pose initialization is severely incorrect (for example, a misaligned
limb or large missing scan region), the Chamfer distance minimization can
converge to a locally consistent but globally incorrect solution. Methods with
stronger statistical shape and pose priors may be less sensitive to poor
initialization, but we did not test this trade-off here. 

\subsubsection{SMPL Model Limitations}
SMPL is learned from surface scans and is not designed to represent
contact-induced soft-tissue deformation. This limitation is relevant for
supine scans, such as MRI-derived body surfaces, where contact with a flat
support can flatten the posterior body surface. Such deformation is not explicitly modeled by SMPL and may limit registration
accuracy in regions affected by sustained contact. Similar issues
can occur in seated configurations, where the supporting surface changes local
soft-tissue geometry. 

SMPL joint centers are regressed from surface shape rather than measured
from internal anatomy. Joint positions recovered by a forward pass on the
registered mesh are therefore biomechanically incorrect and should be used
with care in kinematic analysis. 

The anthropometric post-processing tool also inherits limitations from
SMPL segmentation. SMPL segment boundaries are defined by skinning weights
rather than anatomical landmarks, so segment volumes and cross-sectional areas
near boundaries can differ from measurements based on anatomical definitions.
The reported measurements should therefore be treated as outputs of a
consistent SMPL-based segmentation, not as direct anatomical
measurements. 
\subsection{Future Work}
\label{sec:discussion_future}
During visual inspection, we observed local surface deviations around the
breast, where soft-tissue shape varies substantially across participants. The
standard SMPL segmentation groups this surface within broader trunk segments.
Therefore, the segment-level errors reported in
Table~\ref{tab:combined_segment_accuracy} cannot quantify this local behavior
directly.

Future work could use finer segmentations than the existing 
SMPL segments when local soft-tissue deformation is of interest, for example by
defining a separate breast region. Another direction is to develop body models with region-adaptive mesh detail in soft-tissue regions, so that localized deformations are represented with higher
spatial resolution and measured within matching anatomical segments. 

\section*{Compliance with Ethical Standards}
The MorphoMotion dataset was collected under approval by the Human Research Ethics
Committee of Delft University of Technology (application IDs: 3718, 4076, 4268).
All participants provided written informed consent.

\section*{CRediT Authorship Contribution Statement}
Chaurasia V.: Conceptualization, Methodology, Software, Formal analysis,
Writing -- original draft, Writing -- review and editing.
Cueto Fernandez J.: Data acquisition, Writing -- review and editing.
Prendergast Micah J.: Conceptualization, Methodology,  Writing -- review and editing, Funding acquisition.
van der Kruk E.: Conceptualization, Methodology, Writing -- review and editing, Resources,  Funding acquisition.

\section*{Declaration of Competing Interest}
The authors declare that they have no known competing financial interests or personal relationships that could have appeared to influence the work reported in this paper.

\section*{Acknowledgments}
This work was supported by the European Research Council through the ERC Starting Grant Diversity Outside In (grant number 101221158). We thank the CHI3D dataset creators for providing the data and documentation used in this study. We thank the developers of the SMPL models for providing access. We also thank the participants who took part in the collection of the MorphoMotion dataset. 
\appendix 
\section{MediaPipe-to-SMPL Keypoint Conversion}
\label{sec:appendix_keypoint_mapping}
\begin{figure}[ht!]
    \centering
   \begin{overpic}[width=0.5\linewidth,  unit=2bp,tics=7 ] {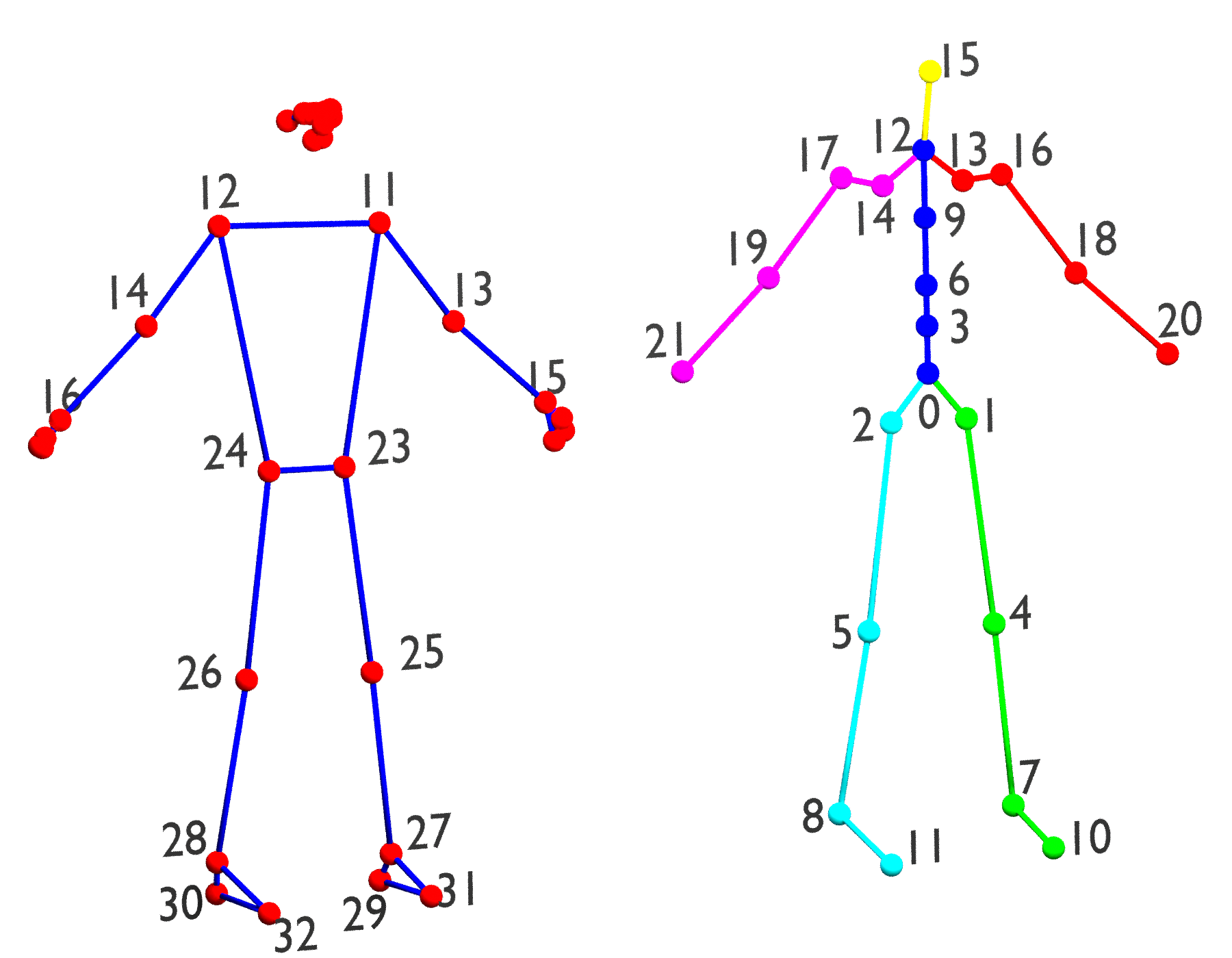}
    \end{overpic}
\caption{MediaPipe and SMPL keypoint skeletons used in pose initialization. MediaPipe landmarks are shown on the left and the converted SMPL keypoints on the right. A few labels in the MediaPipe skeleton are omitted for clarity. The SMPL skeleton is used to initialize the template prior to shape and pose optimization.}
\label{fig:mediapipe_smpl_mapping}
\end{figure}
We use MediaPipe detections only as an intermediate representation for
pose initialization. The optimization itself is performed using SMPL-family
joints. Because the MediaPipe and SMPL skeletons use different joint
definitions, we convert the detected MediaPipe landmarks into a 22-joint SMPL
keypoint set before fitting the initial pose. In Figure~\ref{fig:mediapipe_smpl_mapping}, we show the MediaPipe landmark skeleton and the converted SMPL keypoint skeleton used by BODIESReg. 
MediaPipe provides 33 landmarks in each projected view. We combine detections from the frontal and sagittal projections to obtain 3D landmark coordinates and visibility scores. The frontal projection provides the $X$ and $Z$ coordinates, while the sagittal projection provides the $Y$ and $Z$ coordinates. When a landmark is visible in both projections, the two visibility scores are combined and used as a confidence measure.
 We then map these MediaPipe landmarks to the 22 SMPL joints used for
pose initialization. Joints with direct anatomical counterparts are copied from
the corresponding MediaPipe landmark, and the remaining joints are derived from
neighboring landmarks. In Table~\ref{tab:mediapipe_smpl_mapping}, we list the
MediaPipe landmarks used for each SMPL joint and indicate whether each joint is
copied directly or computed from neighboring landmarks. 
\begin{table}[!ht]
\centering
\caption{MediaPipe-to-SMPL keypoint mapping used for pose initialization.}
\vspace{0.2in}
\label{tab:mediapipe_smpl_mapping}
\renewcommand{\arraystretch}{1.08}
\footnotesize
\begin{tabular*}{0.5\textwidth}{@{}clcl@{}}
\toprule
\textbf{SMPL} & \textbf{Joint} & \textbf{MediaPipe} & \textbf{Rule} \\
\midrule
0  & Pelvis         & 23,24      & hip midpoint  \\
1  & Left hip       & 23         & direct \\
2  & Right hip      & 24         & direct \\
3  & Spine1         & 0,12       & pelvis--neck interp. \\
4  & Left knee      & 25         & direct \\
5  & Right knee     & 26         & direct \\
6  & Spine2         & 0,12       & pelvis--neck interp. \\
7  & Left ankle     & 27         & direct \\
8  & Right ankle    & 28         & direct \\
9  & Spine3         & 0,12       & pelvis--neck interp. \\
10 & Left foot      & 31         & direct \\
11 & Right foot     & 32         & direct \\
12 & Neck           & 11,12,15   & shoulders--head \\
13 & Left collar    & 12,16      & neck--shoulder \\
14 & Right collar   & 12,17      & neck--shoulder \\
15 & Head           & 0,2,5,7,8  & face landmarks \\
16 & Left shoulder  & 11         & direct \\
17 & Right shoulder & 12         & direct \\
18 & Left elbow     & 13         & direct \\
19 & Right elbow    & 14         & direct \\
20 & Left wrist     & 15         & direct \\
21 & Right wrist    & 16         & direct \\
\bottomrule
\end{tabular*}
\end{table}
 For joints without a direct MediaPipe counterpart in
Table~\ref{tab:mediapipe_smpl_mapping}, we compute their locations from
neighboring landmarks. The pelvis is first
estimated as the midpoint of the left and right hips. The head is estimated
from face landmarks: when both ears and both eyes are available, we use the
midpoint of the ears and shift it along the eye-to-ear direction; when only
ears and nose are available, we use the nose-to-ear direction; when only the
nose is available, the nose is used as a fallback. The neck is placed between
the shoulder midpoint and the estimated head. The collar joints are placed
between the neck and the corresponding shoulder. The spine joints are
interpolated along the pelvis-to-neck line, and the pelvis and collar positions
are then adjusted consistently with this estimated spine. 
Each SMPL keypoint is assigned a confidence score derived from the MediaPipe visibility scores of the source landmarks. Direct correspondences inherit the combined MediaPipe visibility. Derived joints use averaged or scaled confidences from their source landmarks. We also record whether each SMPL keypoint was directly detected or derived, so the pose-initialization stage can use one consistent SMPL-format landmark set while retaining detection-quality information.
 \section{Correction of Left--Right Landmark Assignments}
\label{sec:keypoint_left_right_correction}
We use the coordinate convention defined in Section~\ref{sec:coord_alignment}, where the $X$-axis is lateral, the $Y$-axis is anterior--posterior, and the $Z$-axis is vertical. The aligned subject is expected to face the positive $Y$-direction. Under this convention, landmarks on the left side of an upright subject have smaller $X$-coordinates than their corresponding right-side landmarks. The opposite ordering applies when the subject is upside down or faces the negative $Y$-direction.

MediaPipe left--right landmark labels can be inconsistent in near-symmetric poses. We therefore inspect the reconstructed 3D landmarks before mapping them to the SMPL joints. This correction requires the nose, both shoulders, and both hips. If any of these landmarks are unavailable, we retain the original assignments.

Let $\bfp_{\mathrm{n}}$ denote the nose position, $\bfp_{\mathrm{ls}}$ and $\bfp_{\mathrm{rs}}$ the left and right shoulder positions, and $\bfp_{\mathrm{lh}}$ and $\bfp_{\mathrm{rh}}$ the left and right hip positions. We calculate the shoulder and hip midpoints as
\be
\bfm_{\mathrm{s}}= \frac{\bfp_{\mathrm{ls}}+\bfp_{\mathrm{rs}}}{2} \quad \text{and} \quad
\qquad \bfm_{\mathrm{h}}=\frac{\bfp_{\mathrm{lh}}+\bfp_{\mathrm{rh}}}{2},
\ee
respectively.
We classify the subject as upright when the $Z$-coordinate of $\bfm_{\mathrm{s}}$ is greater than that of $\bfm_{\mathrm{h}}$. We estimate the facing direction from the $Y$-component of the vector from the shoulder midpoint to the nose,
\be
d_{\mathrm{n}}= [\bfp_{\mathrm{n}}-\bfm_{\mathrm{s}}]_y.
\ee
For $d_{\mathrm{n}}>0$, the subject faces the positive $Y$-direction, and for $d_{\mathrm{n}}<0$, the subject faces the negative $Y$-direction. We then examine the lateral ordering of the shoulders and hips using
\be
d_{\mathrm{s}}
=
\left[\bfp_{\mathrm{ls}}-\bfp_{\mathrm{rs}}\right]_x,
\qquad
d_{\mathrm{h}}
=
\left[\bfp_{\mathrm{lh}}-\bfp_{\mathrm{rh}}\right]_x.
\ee
 \begin{table}[t]
\centering
\renewcommand{\arraystretch}{1.35}
\begin{tabular}{@{}c@{\hskip 2.85em}c@{\hskip 2.85em}c@{}}
\hline
Facing & Body & Expected \\
direction & orientation & ordering \\
\hline
$d_{\mathrm{n}}>0$ & upright & $d_{\mathrm{s}}<0,\ d_{\mathrm{h}}<0$ \\
$d_{\mathrm{n}}>0$ & upside down & $d_{\mathrm{s}}>0,\ d_{\mathrm{h}}>0$ \\
$d_{\mathrm{n}}<0$ & upright & $d_{\mathrm{s}}>0,\ d_{\mathrm{h}}>0$ \\
$d_{\mathrm{n}}<0$ & upside down & $d_{\mathrm{s}}<0,\ d_{\mathrm{h}}<0$ \\
\hline
\end{tabular}
\vspace{0.2in}
\caption{Expected lateral ordering for the left--right consistency check.}
\label{tab:left_right_correction}
\end{table}
We apply the left--right correction whenever the shoulder ordering is inconsistent with Table~\ref{tab:left_right_correction}. If the shoulder ordering is consistent but the hip ordering is not, we inspect the knee ordering when both knees are available. We apply the correction in this case only when the knee ordering supports the hip ordering.
When correction is required, we exchange the reconstructed coordinates and combined visibility score of each available left--right landmark pair. These pairs comprise the eyes, ears, mouth corners, shoulders, elbows, wrists, fingers, hips, knees, ankles, heels, and foot indices. This operation changes the anatomical assignments without rotating or otherwise transforming the scan.
We subsequently map the corrected landmarks to the $D_k$ SMPL joint locations used for pose initialization. The heuristic can remain ambiguous when required landmarks are missing, their detections are inaccurate, or the torso is strongly rotated relative to the aligned coordinate axes. In these cases, users can correct the joint locations using the interactive pose editor in manual mode.
\section{Effect of Pose-Aligned Initialization: Ablation Study}
\label{sec:ablation}
We next isolate the effect of pose-aligned initialization on registration success. In the ablated condition, we omitted the inverse-kinematics optimization in \eqref{eq:pose_estimation_loss} and set $\bfbeta_0=\bfo$ and $\bftheta_0=\bfo$, corresponding to the default mean shape and T-pose of the SMPL template. These default parameters replaced the inverse-kinematics estimates as both the initial values and regularization anchors in \eqref{eq:chamfer_opt}. Scan preprocessing, model variants, Chamfer optimization, correspondence refinement, regularization weights, iteration counts, and failure assessment were otherwise identical to those used for the full pipeline. 

\begin{figure}[ht!]
    \centering
   \begin{overpic}[width=0.5\linewidth,  unit=2bp,tics=7 ] {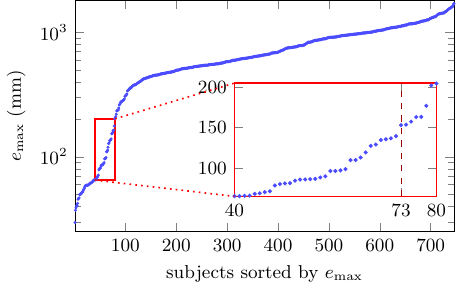}
    \end{overpic}
\caption{Maximum joint error $e_{\max}$, defined in \eqref{eq:max_joint_error}, for all 746 CHI3D registrations
performed without pose initialization, sorted in ascending order. The inset
magnifies the transition region around the failure boundary. Because the error
grows gradually rather than through a sharp jump, the boundary was located by
manual inspection: registrations up to index 72 were verified to be
anatomically correct, while those beyond it exhibit incorrect poses.}
\label{fig:chi3d_Apose}
\end{figure}
\begin{figure}[ht!]
    \centering
   \begin{overpic}[width=0.7\linewidth,  unit=2bp,tics=7 ] {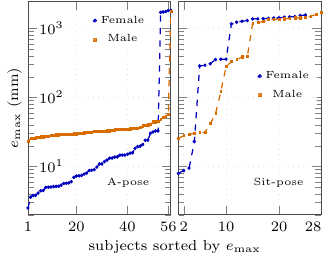}
    \end{overpic}
\caption{Maximum joint error $e_{\max}$, defined in \eqref{eq:max_joint_error}, relative to full-pipeline reference joints, for MorphoMotion registrations performed without pose initialization, sorted in increasing order for each pose and sex. Left: A-pose registrations exhibit an abrupt transition between correct and failed registrations. Right: sit-pose registrations exhibit a more gradual transition. Intermediate errors of $121$--$390$~mm correspond primarily to incorrect hand locations on the thighs or abdomen, whereas errors of order $10^3$~mm correspond to registrations with an approximately $180$-degree body-orientation flip. We determined the sit-pose failure boundary by visual inspection.} 
\label{fig:morpho_Apose}
\end{figure}
For the CHI3D dataset, the full pipeline registered 618 of 746 scans ($82.9\%$) correctly, whereas registration from the default T-pose registered only 72 scans ($9.7\%$) correctly; the remaining 674 registrations ($90.3\%$) failed according to the criterion defined in Subsection~\ref{sec:joint_failure_criterion}. In Figure~\ref{fig:chi3d_Apose}, we report the maximum joint error $e_{\max}$ for all 746 ablated registrations, sorted in ascending order.  Because the error increased gradually without a sharp separation between correct and failed registrations, we visually inspected each registration within the transition region. We classified a registration as correct when the head, torso, arms, hands, legs, and feet were aligned with their corresponding anatomical regions; any visibly misplaced body part was classified as a failure. We placed the failure boundary at the first anatomically incorrect registration. This inspection confirmed that the first 72 registrations were anatomically correct, whereas the remaining registrations converged to incorrect poses.

Ground-truth joint locations are unavailable for MorphoMotion. We therefore defined reference joints from the full automatic pipeline: each MorphoMotion scan was registered with the full pipeline, these registrations were visually verified as anatomically correct, and the corresponding SMPL parameters were passed through the SMPL forward model to obtain joint locations. For the ablation study, the same scans were registered from the default T-pose, without pose-aligned initialization. We then passed the optimized ablated SMPL parameters through the SMPL forward model to obtain ablated joint locations, and computed the maximum joint error $e_{\max}$, defined in \eqref{eq:max_joint_error}, relative to the full-pipeline reference joints. We used these errors to sort the ablated registrations and locate the transition regions in Figure~\ref{fig:morpho_Apose}. Final classification as correct or failed was based on visual inspection of registrations near each transition region.

The A-pose registrations for both sexes exhibited a sharp transition between correct and failed registrations, as shown in the left panel of Figure~\ref{fig:morpho_Apose}.  
The sit-pose curves for both sexes, shown in the right panel of Figure~\ref{fig:morpho_Apose}, contain two failure regimes. By visual inspection, we found that the intermediate errors of $121$--$390$~mm arose primarily when the registered hands were placed on the thighs or merged with the abdomen. Because this configuration brings the hands close to these surfaces, these anatomically incorrect registrations produce smaller joint errors. The second regime, with errors of order $10^3$~mm, corresponds to registrations with an approximately $180$-degree body-orientation flip. We applied the same visual criteria to registrations in the MorphoMotion transition regions and obtained the following results for registration without pose initialization: 
\begin{center}
\begingroup
\setlength{\tabcolsep}{8pt}
\renewcommand{\arraystretch}{1.25}
\footnotesize
\begin{tabular}{|l|c|c|c|}
\hline
\textbf{Group} & \textbf{Scans} & \textbf{Correct} & \textbf{Failed} \\
\hline
Female A-pose   & 57 & 52 ($91.2\%$) & 5 ($8.8\%$) \\
\hline
Male A-pose     & 57 & 56 ($98.2\%$) & 1 ($1.8\%$) \\
\hline
Female sit-pose & 25 & 4 ($16.0\%$)  & 21 ($84.0\%$) \\
\hline
Male sit-pose   & 28 & 8 ($28.6\%$)  & 20 ($71.4\%$) \\
\hline
\end{tabular}
\endgroup
\end{center}
Across sexes, registration from the default T-pose failed for 6 of 114 A-pose scans ($5.3\%$), but for 41 of 53 sit-pose scans ($77.4\%$). Pose initialization therefore had a larger effect when the scan pose differed substantially from the default T-pose.  
\section{Anthropometric Measurements from Registered SMPL Meshes}
\label{sec:appendix_anthropometrics}
We use the registered SMPL meshes to extract three anthropometric
quantities: segmental volumes, cross-sectional area profiles, and segment
lengths. We present these measurements as illustrative post-processing outputs
from registered SMPL meshes, not as independently validated anthropometric
measurements.

These quantities are commonly used in inertial parameter estimation, body
composition analysis, and clinical anthropometry. Each vertex in the registered
SMPL mesh inherits an anatomical label from the template, which partitions the
registered mesh into 24 body segments, as shown in
Figure~\ref{fig:cross_section}. Because the labels come directly from the
registered template, no separate anatomical labeling step is required. 
\subsection{Segmental Volume}
\label{sec:volume}
We estimate the volume of each body segment from the registered vertices assigned to that segment. The post-processing tool applies an unconstrained Delaunay tetrahedralization to these vertices and sums the volumes of the resulting tetrahedra. For a tetrahedron with vertices $\mathbf{p}_1, \mathbf{p}_2, \mathbf{p}_3, \mathbf{p}_4$, the volume is
\begin{equation}
V_{\text{tet}} = \frac{1}{6} \bigl|\bigl((\mathbf{p}_2 - \mathbf{p}_1) \times (\mathbf{p}_3 - \mathbf{p}_1)\bigr) \cdot (\mathbf{p}_4 - \mathbf{p}_1)\bigr|,
\end{equation}
and the segmental volume estimate is
\begin{equation}
V_{\text{seg}} = \sum_{k} V_{\text{tet},k},
\end{equation}
where the sum is taken over all tetrahedra returned by the tetrahedralization. Because the tetrahedralization is computed from segment vertices without enforcing the anatomical surface as a constrained boundary, $V_{\text{seg}}$ should be interpreted as a vertex-enclosure volume estimate rather than a strictly surface-bounded anatomical volume. This distinction is most relevant for concave segments or regions whose assigned vertices do not form a closed anatomical boundary. If the SciPy/Qhull tetrahedralization fails because the points are coplanar, nearly degenerate, duplicated, too sparse, or numerically ill-conditioned, we use the convex hull volume as a fallback approximation. Segments with fewer than four vertices are assigned zero volume.
\subsection{Cross-Sectional Area Profile}
\label{sec:cross_section}
The cross-sectional area at a position along a segment quantifies local body geometry and is the input for computing segment mass distributions used in inertial parameter models. Given the proximal and distal joint positions $Q_1$ and $Q_2$ of a segment, the segment axis is
\begin{equation}
\hat{\mathbf{d}} = \frac{Q_2 - Q_1}{\|Q_2 - Q_1\|}.
\end{equation}
The cutting plane at fractional position $t \in [0,1]$ along the segment is centered at
\begin{equation}
\mathbf{c}(t) = Q_1 + t\,\|Q_2 - Q_1\|\,\hat{\mathbf{d}},
\end{equation}
with normal $\hat{\mathbf{d}}$. We select mesh vertices within a thin slab of half-thickness $\varepsilon$ around this plane and project them onto it using two tangent vectors $\hat{\mathbf{t}}_1$, $\hat{\mathbf{t}}_2$ that form an orthonormal basis with $\hat{\mathbf{d}}$:
\begin{equation}
\mathbf{v}_{2D} = \bigl(({\mathbf{v} - \mathbf{c}}) \cdot \hat{\mathbf{t}}_1,\;(\mathbf{v} - \mathbf{c}) \cdot \hat{\mathbf{t}}_2\bigr).
\end{equation}
The cross-sectional area is the convex hull area of the projected points, computed by the shoelace formula:
\begin{equation} \label{eq:At}
A(t) = \frac{1}{2}\left|\sum_{i=1}^{n} (x_i y_{i+1} - x_{i+1} y_i)\right|,
\end{equation}
where $(x_i, y_i)$ are the projected vertices ordered counterclockwise and indices are taken modulo $n$. We sweep $t$ from 0 to 1 to obtain the full cross-sectional area profile of the segment.

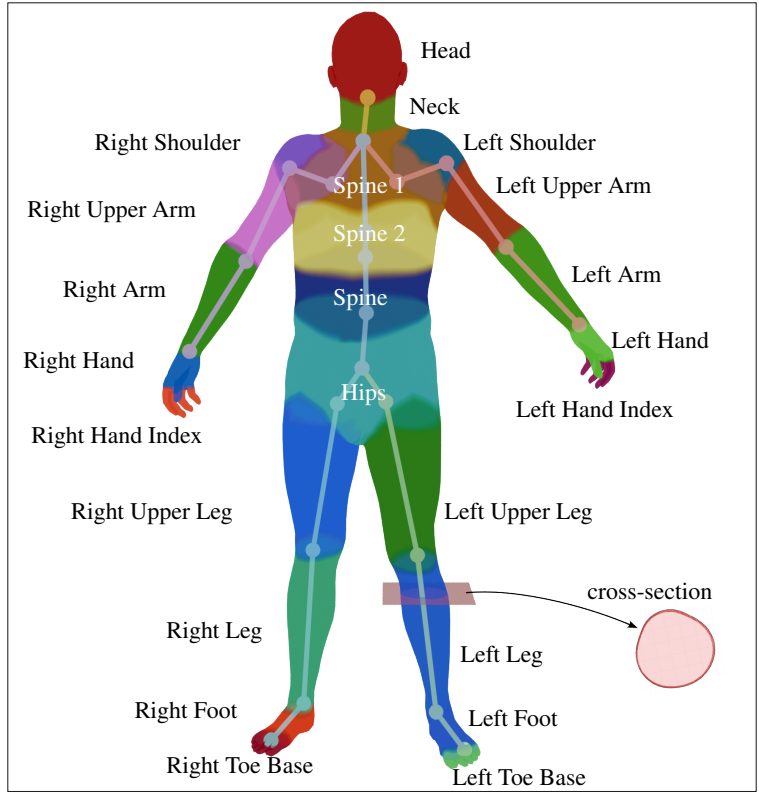
\begin{figure}[ht!]
\centering
   \begin{overpic}[width=0.6\linewidth,  unit=1bp,tics=7  ] {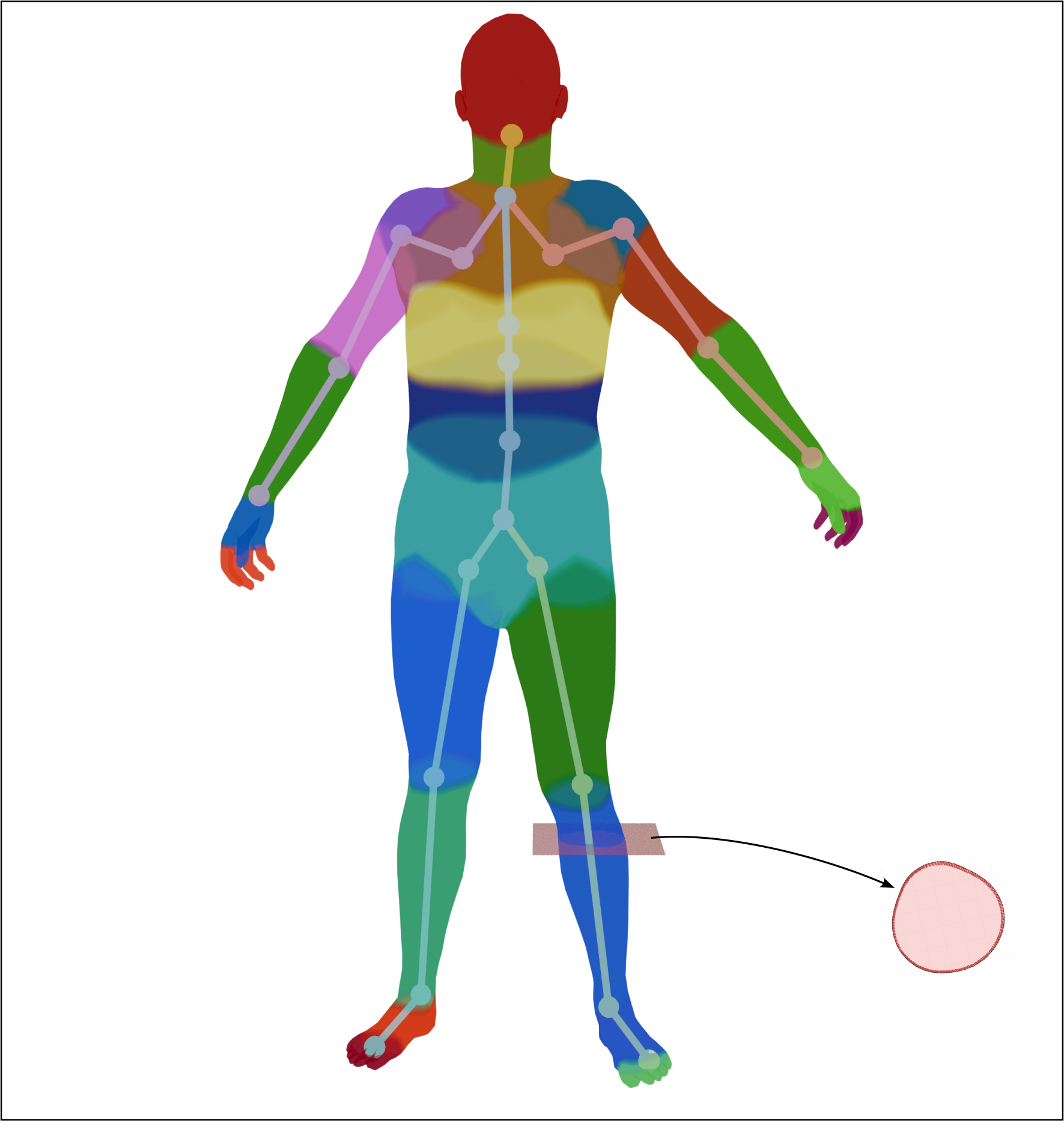}
 \put(52,93){\footnotesize Head}
  \put(50.6,86){\footnotesize Neck}
   \put(57.4,81.5){\footnotesize Left Shoulder}
    \put(11,81.5){\footnotesize Right Shoulder}
     \put(61.5,76){\footnotesize Left Upper Arm}
      \put(2.5,73){\footnotesize Right Upper Arm}
       \put(71,64.9){\footnotesize Left Arm}
        \put(7,63){\footnotesize Right Arm}
         \put(76,56.5){\footnotesize Left Hand}
        \put(2,54){\footnotesize Right Hand}
         \put(64,47.9){\footnotesize Left Hand Index}
          \put(3,44.6){\footnotesize Right Hand Index}
           \put(42,50){\footnotesize \clw Hips}
            \put(41,76){\footnotesize \clw Spine 1}
             \put(41,70){\footnotesize \clw Spine 2}
              \put(41,62){\footnotesize \clw Spine}
               \put(55,35){\footnotesize  Left Upper Leg}
                \put(8,35){\footnotesize   Right Upper Leg}
                 \put(57,17){\footnotesize  Left Leg}
                  \put(20,20){\footnotesize  Right Leg}
                   \put(58,9){\footnotesize Left Foot}
                    \put(16,10){\footnotesize Right Foot}
                      \put(56,1.5){\footnotesize Left Toe Base}
                       \put(20,3){\footnotesize Right Toe Base}
                        \put(73,25){\footnotesize  cross-section}
     \end{overpic}
\caption{SMPL template segmented into 24 anatomical regions, each shown in a distinct color, with the underlying skeleton of 22 joints and 21 bones. The inset shows a representative cross-section at a selected body location: vertices near the cutting plane and the convex hull fitted to those vertices to estimate cross-sectional area.}
        \label{fig:cross_section}
\end{figure}

The cutting-plane location can be specified in two ways. First, as a fractional position $t \in [0,1]$ where $t=0$ is the proximal joint and $t=1$ is the distal joint. Second, as an absolute distance $d$ from either joint, which is converted internally to $t = d / \|Q_2 - Q_1\|$. Both modes use the same shoelace formula in \eqref{eq:At}. In Figure~\ref{fig:cross_section}, we show the 24-region SMPL segmentation and underlying skeleton, with an example cross-section used to estimate local cross-sectional area.
\subsection{Segment Lengths}
\label{sec:seg_lengths}
Segment lengths are computed directly from the joint positions of the fitted SMPL model. The length of a segment bounded by joints $Q_1$ and $Q_2$ is $\|Q_2 - Q_1\|$. Because the joint positions are a function of the optimized parameters $(\bfbeta^{**}, \bftheta^{**})$, no separate post-processing is required.  
\subsection{Accuracy and Limitations}
\label{sec:anthropometrics_limitations}
Two limitations bound the accuracy of the anthropometric measurements.
First, SMPL segment boundaries are defined by the skinning topology of the
model, not by biomechanical joint definitions.

The SMPL joint centers are regressed from surface shape, so segment boundaries
at the hip and shoulder do not coincide exactly with the anatomical landmarks
used in clinical anthropometry or musculoskeletal modeling. Volumes and
cross-sectional areas near segment boundaries will therefore differ from
measurements made by a trained anatomist using those landmarks.

Second, we have no ground-truth measurements with which to quantify errors in
cross-sectional area or segmental volume. Surface-fit error influences these
measurements, but the relationship depends on segment dimensions, local
curvature, error direction, and spatial distribution.
Mean vertex distance alone therefore cannot provide quantitative bounds on
area or volume error.  

Therefore, the anthropometric measurements are best suited for comparative analysis between subjects registered to the same pipeline, rather than for absolute clinical measurement. Within that scope, the tools apply a consistent segmentation and measurement procedure without additional manual labeling after registration.
\section*{Declaration of Generative AI and AI-Assisted Technologies in the Manuscript Preparation Process}
During the preparation of this work, the authors used ChatGPT (OpenAI) for language polishing and proofreading. The authors reviewed and edited the output as needed and take full responsibility for the content of the published article.

\end{document}